\documentclass[sn-mathphys-ay]{sn-jnl}

\usepackage{graphicx}%
\usepackage{multirow}%
\usepackage{amsmath,amssymb,amsfonts}%
\usepackage{amsthm}%
\usepackage{mathrsfs}%
\usepackage[title]{appendix}%
\usepackage{xcolor}%
\usepackage{textcomp}%
\usepackage{manyfoot}%
\usepackage{booktabs}%
\usepackage{algorithm}%
\usepackage{algorithmicx}%
\usepackage{algpseudocode}%
\usepackage{listings}%
\usepackage{subcaption}
\usepackage{mdframed}
\usepackage[normalem]{ulem}
\usepackage{caption}

\newcommand*\diff{\mathop{}\!\mathrm{d}}

\definecolor{kellygreen}{rgb}{0.2, 0.6, 0}

\definecolor{changesColour}{rgb}{0.75, 0, 0.7}
\newcommand\changes[1]{#1} 

\newcommand\chaser{c}
\newcommand\Chaser{C}
\newcommand\runner{\rho}
\newcommand\Runner{P}

\let\olditemize\itemize \renewcommand{\itemize}{\olditemize\itemsep10pt}

\definecolor{BoxGreen}{RGB}{213, 241, 229}
\newmdenv[
  linewidth=0pt, 
  linecolor=white, 
  innerleftmargin=10pt, 
  innerrightmargin=10pt, 
  innertopmargin=7pt, 
  innerbottommargin=7pt, 
  backgroundcolor=BoxGreen, 
]{boxx}

\usepackage[left]{lineno}  

\begin{document}

\title[Chase-and-Run and Chirality in Nonlocal Models of Pattern Formation]{Chase-and-Run and Chirality in Nonlocal Models of Pattern Formation}

\author*[1]{\fnm{Thomas Jun} \sur{Jewell}}\email{jewell@maths.ox.ac.uk}

\author[2]{\fnm{Andrew L.} \sur{Krause}}\email{andrew.krause@durham.ac.uk}

\author[1]{\fnm{Philip} \sur{K. Maini}}\email{maini@maths.ox.ac.uk}

\author[1]{\fnm{Eamonn} \sur{A. Gaffney}}\email{gaffney@maths.ox.ac.uk}

\affil*[1]{\orgdiv{Wolfson Centre for Mathematical Biology, Mathematical Institute}, \orgname{University of Oxford}, \orgaddress{Andrew Wiles Building, \street{Radcliffe Observatory Quarter, Woodstock Road}, \city{Oxford}, \postcode{OX2 6GG}, \country{United Kingdom}}}

\affil[2]{\orgdiv{Mathematical Sciences Department}, \orgname{Durham University}, Upper Mountjoy Campus \orgaddress{\street{Stockton Rd}, \city{Durham}, \postcode{DH1 3LE}, \country{United Kingdom}}}

\abstract{Chase-and-run dynamics, in which one population pursues another that flees from it, are found throughout nature, from predator-prey interactions in ecosystems to the collective motion of cells during development. Intriguingly, in many of these systems, the movement is not straight; instead, `runners' veer off at an angle from their pursuers. This angled movement often exhibits a consistent left–right asymmetry, known as lateralisation or chirality. Inspired by such phenomena in zebrafish skin patterns and evasive animal motion, we explore how chirality shapes the emergence of patterns in nonlocal (integro-differential) advection-diffusion models. We extend such models to allow movement at arbitrary angles, uncovering a rich landscape of behaviours. We find that chirality can enhance pattern formation, suppress oscillations, and give rise to entirely new dynamical structures, such as rotating pulses of chasers and runners. We also uncover how chase-and-run dynamics can cause populations to mix or separate. Through linear stability analysis, we identify physical mechanisms that drive some of these effects, whilst also exposing striking limitations of this theory in capturing more complex dynamics. Our findings suggest that chirality could have roles in ecological and cellular patterning beyond simply breaking left-right symmetry.}


\maketitle

\section{Introduction}\label{sec1}
Across biology, organisms and cells alike move in response to others around them, giving rise to striking population-level patterns. From territorial ranges \citep{potts2014_territories} to swarming locusts \citep{bazaziCouzin2008locust_cannibalism} to the development of tissues and organs \citep{hillenPainter_user_guide_to_pde_chemotax, armstrong2006_nonlocal}, these dynamics emerge from simple rules of interaction. Theoretical models can bridge these scales, revealing when and how individual-level behaviours give rise to macroscopic structure. Yet our understanding is far from complete. Even between just two species, there is a wide variety of possible interaction dynamics.

For example, movement-inducing interactions between different species can be mutually attracting, mutually avoidant, or `chase-and-run', in which individuals of the first species are attracted towards individuals of the second species, whilst individuals of the second species are repelled away from individuals of the first. Chase-and-run dynamics, also called `pursue-and-avoid', `cops-and-robbers', or `chase-and-escape', are often studied in the context of ecosystems with predators and prey \citep{moore2015_chase_run_ecology_review}. One example includes North American elk, which have been observed to alter their browsing patterns and spatial distribution to avoid wolves \citep{white1998_elk_wolves, ripple2000_elk_wolves}. Over the last two decades, ecologists have adopted the idea that predator-avoidant movement in prey and prey-seeking movement in predators can drive spatial patterning across a landscape \citep{gaynor2019landscape_of_fear}. Additionally, chase-and-run dynamics can occur at cellular scales, where they are also thought to contribute to pattern formation. Examples include stripe formation in zebrafish \citep{YamanakaKondo2014vitro} and cell migration for neural crest development \citep{theveneau2013_chase_run_neural_crest}. \citet{szabo2015_chase_run_common} argue that chase-and-run may be a common mechanism present in many cellular pattern forming processes, possibly including kidney morphogenesis, development of the lateral line in zebrafish, and cancer invasion.

In response to this growing interest, \citet{painter_giunta_2024_chase_run} recently studied chase-and-run interactions within a nonlocal advection-diffusion model. They showed that these interactions can generate a rich range of behaviour, including stationary patterns, spatio-temporal oscillations, and `population chase-and-run' -- where an aggregate of chasers continually chases an aggregate of runners. This work is part of a broader field examining nonlocal advection-diffusion models, which use integro-partial differential equations to describe agents that interact across a distance to induce movement. Such models have been applied to various biological phenomena, including swarms of animals, which rely on visual cues across a distance \citep{mogilner1999_nonlocal_swarm}; cell aggregation, where adhesion occurs across the length of a cell diameter and longer ranged interactions can be mediated by pseudopodia \citep{armstrong2006_nonlocal, carrillo2019_nonlocal_adhesion, Painter2015nonlocal, jewell2023}; and territorial patterns formed by animals communicating through scent markings across an ecosystem \citep{potts2019_nonlocal_memory}. For a recent review of these models, see \citet{painter_hillen_potts_nonlocal_review}.

An assumption present in all the models mentioned above is that agents only move directly towards or directly away from each other -- advection is only induced parallel to their separation or parallel to their concentration gradient. What happens when we relax this assumption? What happens when interactions generate advection at some general angle from the source of the signal? These questions are relevant to chase-and-run mechanisms within both ecology and developmental biology. 

In animal behaviour, escape responses often involve movement at an angle rather than directly away from a threat. In many vertebrates, such behaviour can exhibit consistent left–right biases, a phenomenon known as `behavioural lateralisation'. Such asymmetries are well-documented in fish species, both at the individual and population level, and extend beyond escape responses to include foraging, social interactions, and mating behaviours \citep{miletto2020_lateralisation_well_cited_review, gobbo2025_zebrafish_behaviour_lateralisation_mini_review}. For example, \citet{cantalupo1995_population_level_lateralization} observed that individual immature \textit{girardinus falcatus} displayed a preference for escaping either rightwards or leftwards, with rightwards bias being more common. \citet{ghirlanda2004_population_lateralisation_evolution} suggest such population-level bias could offer an evolutionary advantage by improving coordination.  In our work, we will use the term `chirality' to describe consistent left–right asymmetries in movement.

\begin{figure}
    \centering
    \includegraphics[width=0.7\textwidth]{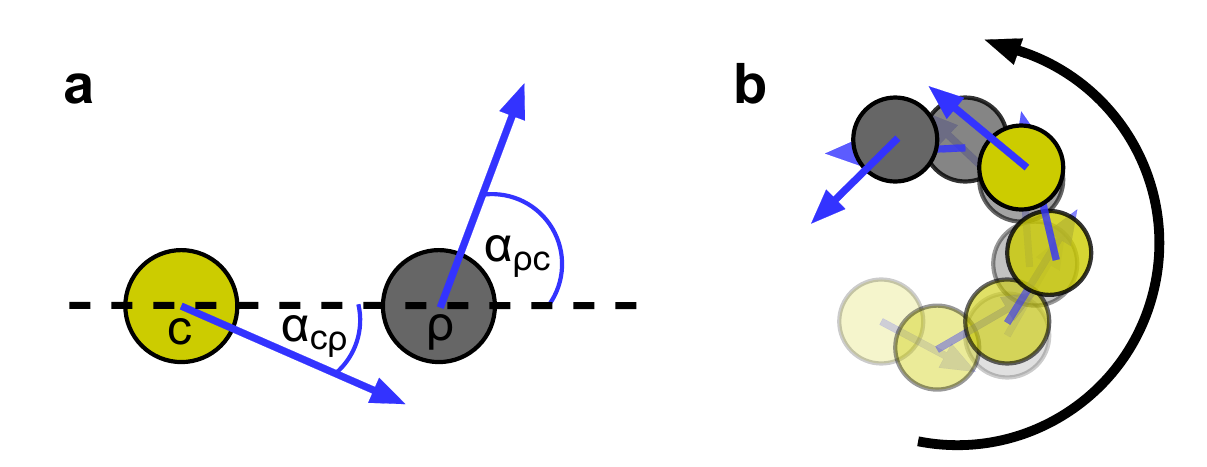}
    \caption{Schematics based on Fig. 3 of \citet{YamanakaKondo2014vitro}. \textbf{a} The mean direction of movement between an interacting xanthophore (yellow pigment cell, labelled $\chaser$) and melanophore (black pigment cell, labelled $\runner$) of wildtype zebrafish \textit{in vitro}. The mean angles observed by \citet{YamanakaKondo2014vitro} were $\alpha_{\chaser\runner}\approx -20^\circ$ and $\alpha_{\runner\chaser}\approx 70^\circ$. \textbf{b} Typical movements resulting from these interactions.}
    \label{fig:chiral_run_chase_diagram}
\end{figure}

A notable example of chiral movement from developmental biology is that of the pigment cells of zebrafish skin. The characteristic stripes of zebrafish are thought to arise from interactions between pigment cells, which migrate in response to contact with pseudopodia from other pigment cells \citep{kondo2021_zebrafish_review}. \textit{In vitro} experiments by \citet{YamanakaKondo2014vitro} showed that melanophores (black pigment cells) move away from xanthophores (yellow pigment cells), whilst xanthophores move towards melanophores. Interestingly, their movements are not parallel to their separation. Melanophores move consistently anticlockwise at an average angle of roughly $70^{\circ}$ relative to xanthophores, whilst xanthophores move clockwise roughly $20^{\circ}$ relative to melanophores. \textbf{Fig.\,\ref{fig:chiral_run_chase_diagram}} illustrates this chiral chase-and-run mechanism. 

The function of chirality in this cellular system and its impact on self-organisation remains an open question. In some biological systems, cell chirality can lead to the emergence of left-right asymmetry in a body plan \citep{nonaka1998_chirality_asymmetricbodyplan, shibazaki2004_chirality_asymmetricbodyplan,taniguchi2011_chirality_asymmetricbodyplan}; however, zebrafish stripes are essentially left-right symmetric \citep{YamanakaKondo2021_chirality_function}. \citet{YamanakaKondo2021_chirality_function} argue that chirality is a property common to many cell types and may have functions beyond creating an asymmetric body plan. They further hypothesise that chirality may contribute to the alignment of parallel zebrafish stripes, although this has not been tested with experiment or modelling. Models by \citet{chen2012_labframe_anisotropy} have shown that anisotropy relative to fixed directions on a domain is sufficient for creating aligned stripes rather than labyrinthine patterns. However, this type of anisotropy is distinct from the chirality of zebrafish cells, which is relative to the direction of separation between each interacting cell, rather than relative to a fixed direction along the medium on which the cells live.  Whilst sophisticated agent based models by \citet{volkening2018iridophores_ABM, owen2020YatesZebrafishABM, cleveland2023_zebrafish_ABM}; and \citet{volkening2015_zebrafish_ABM} can precisely reproduce patterns seen in wildtype and mutant zebrafish, none of these models incorporate chirality in cell movement.

With all this is mind, we will generalise nonlocal advection-diffusion models to allow for chiral movement-interactions and, focusing on chase-and-run dynamics, study how chirality affects pattern forming behaviour. Rather than considering a specific biological system, we take a general approach, with a stylised model, and aim to gain insight on the possible roles of chirality and chase-and-run across biology.

Although other models of agents with chiral movement exist, such as those by \citet{chiralmodel_kreienkamp2022,chiralmodel_kruk2021,chiralmodel_kruk2020,chiralmodel_liebchen2017,chiralmodel_liebchen2022}, these models require the definition of an orientation or polarisation coordinate, usually looking at self-propulsion rather than interaction, and they are often based on Vicsek-type models, which are significantly different to our model. An exception to this is \citet{woolley2017chiral}, who studies a local PDE model derived as a limiting case of a chiral integro-PDE model. He shows that chirality can significantly impact the capacity for pattern formation and give rise to patterns not seen in non-chiral models. Our work naturally extends this approach by looking at fully nonlocal models.

We begin by formulating the models in Section\,\ref{sec:model_formulation}. In Section\,\ref{sec:linear_analysis_pattern_formation}, we conduct a linear stability analysis to determine when patterns can autonomously form from a homogeneous state, and identify where chirality alters these conditions. Section \ref{sec:geometric_argument} provides an intuitive geometric explanation for these results. We then interpret the results in Section\,\ref{sec:insights_PF}, highlighting several key insights on the interplay between chirality and chase-and-run dynamics in driving pattern formation and influencing oscillations. These insights are tested and extended through numerical simulations in Section\,\ref{sec:simulations_1}, where we also examine nonlinear phenomena. We then turn to investigate the impact of chase-and-run interactions on how two species are co-located. Using linear analysis, Section\,\ref{sec:linear_analysis_separation_mixing} derives simple predictive criteria for species mixing or separation. Section \ref{sec:simulations_separation_mixing} uses numerical simulations to highlight where these predictions succeed, and also cases where they breakdown, exploring the nonlinear processes behind these breakdowns. In Section \ref{sec:other_complex_behaviour} we showcase additional emergent phenomena, including population chase-and-run and travelling holes, and explore how these are affected by chirality. Finally, in Section \ref{sec:discussion}, we discuss the implications of our findings.

\section{Model formulation}
\label{sec:model_formulation}
\subsection{Single-species without chirality}
To build intuition for our full two-species chiral models, we begin with a simpler case: nonlocal advection-diffusion models with only a single species and without chirality. Here, a non-dimensional model for a population with density $\runner(\boldsymbol{x},t)$ at position $\boldsymbol{x}\in \mathbb{R}^2$ and at time $t\in [0,\infty)$ is governed by
\begin{equation}
    \frac{\partial \runner }{\partial t} = \nabla^2 \runner  -\boldsymbol{\nabla}\cdot\Biggr( \runner \,\changes{\phi}(\runner )\frac{\mu}{\xi^2} \int_{\mathbb{R}^2}\boldsymbol{\hat{s}}\,\Omega\left(\frac{s}{\xi}\right)\changes{g}(\runner (\boldsymbol{x}+\boldsymbol{s},t))\diff^2\boldsymbol{s}\Biggr),
\label{eq:single_species_direct}
\end{equation}
with the Laplacian term representing diffusion, and the second term on the right-hand side representing advection driven by nonlocal interactions within the population. The population at $\boldsymbol{x}$ interacts with those at each point $\boldsymbol{x}+\boldsymbol{s}$. Here, $\boldsymbol{\hat{s}}$ is the unit vector in the $\boldsymbol{s}$ direction. \changes{For generality, instead of assuming that the advection is simply directly proportional to the density, $\runner(\boldsymbol{x}+\boldsymbol{s},t)$, we allow it to vary as some non-negative, non-decreasing function $g(\runner)$.} The magnitude of interaction also depends on the separation distance, $s\equiv|\boldsymbol{s}|$, through the function $\Omega(\cdot)$. This function, called the `interaction kernel', is typically non-increasing and non-negative, meaning that interactions get weaker with distance and do not change direction with distance. The kernel is normalised such that the interaction strength is parameterised solely by $\mu\in\mathbb{R}$, with $\mu>0$ encoding attraction and $\mu<0$ encoding repulsion. The characteristic interaction range is parameterised by $\xi\in\mathbb{R}_{>0}$. We include a saturation function, $\changes{\phi}(\rho)$, which is non-negative and decreases with density, representing impeded movement due to volume-filling. 

Our analysis below focuses on infinite domains, although imposing a finite periodic domain would simply restrict the allowed pattern wavenumbers. In all cases, we assume the interaction kernel decays sufficiently at large distances to ensure integrals converge on an infinite domain and that wrap-around effects remain negligible on a periodic domain.

Another commonly studied type of nonlocal advection-diffusion model is given by the non-dimensional equation
\begin{equation}
    \frac{\partial \runner }{\partial t} = \nabla^2 \runner  -\boldsymbol{\nabla}\cdot\Biggr( \runner \,\changes{\phi}(\runner )\frac{\mu}{\xi} \boldsymbol{\nabla}\int_{\mathbb{R}^2}\,\changes{\tilde{\Omega}}\left(\frac{s}{\xi}\right)\changes{g}(\runner (\boldsymbol{x}+\boldsymbol{s},t))\diff^2\boldsymbol{s} \Biggr),
\label{eq:single_species_gradient}
\end{equation}
\changes{where $\tilde{\Omega}$ is the interaction kernel for this model.} We refer to Equation \eqref{eq:single_species_direct} as a `direct sensing' model, and Equation \eqref{eq:single_species_gradient} as a `gradient sensing' model. In direct sensing models, advection is induced directly in response to the presence of each individual, and hence is directed parallel to the separation of the individuals, i.e., in the direction of the unit vector $\boldsymbol{\hat{s}}$. The net advection is then the vector sum (integral) of these interactions. In contrast, for gradient sensing models, advection is induced in response to the gradient of a nonlocal measure of the population density. 

Broadly speaking, a direct sensing framework is more often used when interactions involve one-to-one contact, such as for membrane adhesion between cells \citep{Painter2015nonlocal, jewell2023}. The gradient sensing framework is more commonly used to study animal swarming behaviours \citep{mogilner1999_nonlocal_swarm, EdGreen_locust_swarming}. However, both models overlap in application and qualitative behaviour, with both types being used to model cellular and ecological systems \citep{armstrong2006_nonlocal, painter_giunta_2024_chase_run, potts2024_heterogeneous, potts2019_nonlocal_memory, carrillo2019_nonlocal_adhesion}. \changes{In fact, if $\Omega = \frac{\diff}{\diff s}\tilde{\Omega}$, then there is a direct equivalence between these frameworks, which can be seen by use of the divergence theorem, given suitable conditions at any domain boundary \citep{painter_hillen_potts_nonlocal_review}}. Nevertheless, in line with \citet{painter_hillen_potts_nonlocal_review}, which has a more detailed discussion of these models, we will consider them separately, as they have quite different phenomenological interpretations. We will show that our results apply in both frameworks.

Consistent with all of the work cited above, we assume an infinite domain for our analysis and a finite domain with periodic boundary conditions for numerical simulations. Extending such models to other boundary conditions is technically intricate and beyond the scope of this work, although some progress has been made by \citet{new_operators_no_flux, 1D_no_flux_existence, 2D_no_flux_existence}. Another common assumption that we inherit is that the system is mass conserving, as the dynamics are driven purely by transport, rather than proliferation and death.

\subsection{Two-species with chirality}
We generalise Equations \eqref{eq:single_species_direct} and \eqref{eq:single_species_gradient} by introducing chirality, allowing advection to be induced at any angle, $\alpha$, relative to the separation of individuals or the concentration gradient. This is achieved using a rotation matrix, $\boldsymbol{\underline{\underline{R}}}(\alpha)$. For direct sensing models, we replace $\boldsymbol{\hat{s}}$ with the rotated vector
\begin{equation}
    \boldsymbol{\underline{\underline{R}}}(\alpha)\boldsymbol{\hat{s}} \equiv \begin{pmatrix} \text{cos}\,\alpha & -\text{sin}\,\alpha\\\text{sin}\,\alpha & \text{cos}\,\alpha\end{pmatrix}\boldsymbol{\hat{s}}\,.
\label{eq:rotation_matrix_direct}
\end{equation}
For gradient sensing models, we rotate the gradient vector of the nonlocal measure of the population such that
\begin{equation}
\boldsymbol{\nabla}\int_{\mathbb{R}^2}\,\changes{\tilde{\Omega}}\left(\frac{s}{\xi}\right)\changes{g}(\runner (\boldsymbol{x}+\boldsymbol{s},t))\diff^2\boldsymbol{s} \longrightarrow \boldsymbol{\underline{\underline{R}}}(\alpha)\left(\boldsymbol{\nabla}\int_{\mathbb{R}^2}\,\changes{\tilde{\Omega}}\left(\frac{s}{\xi}\right)\changes{g}(\runner (\boldsymbol{x}+\boldsymbol{s},t))\diff^2\boldsymbol{s}\right).
\label{eq:rotation_matrix_gradient}
\end{equation}
In both cases, $-90^\circ<\alpha\leq 90^\circ$, so that the distinction between attraction and repulsion is still governed by the sign of the interaction strength, $\mu$.

Finally, to investigate chase-and-run dynamics, we extend the models to two species, incorporating self-interactions between members of the same species and cross-interactions between members of different species. For generality, we assume that each of these four interactions can have its own interaction strength, interaction length, kernel, angle, and volume-filling function. 

In the case of direct sensing, the density of a populations of chasers, $\chaser(\boldsymbol{x},t)$, and the density of a population of runners, $\runner(\boldsymbol{x},t)$, therefore evolves as
\begin{equation}
\begin{split}
    \frac{\partial \chaser}{\partial t} = \nabla^2 \chaser -\boldsymbol{\nabla}\cdot\Biggr( \chaser \changes{\phi}_\chaser (\chaser ,\runner )\Biggr[\frac{\mu_{\chaser \chaser }}{\xi_{\chaser \chaser }^2} \int_{\mathbb{R}^2}  \boldsymbol{\underline{\underline{R}}}(\alpha_{\chaser \chaser })\boldsymbol{\hat{s}}\,\Omega_{\chaser \chaser }\left(\frac{s}{\xi_{\chaser \chaser }}\right)g_{\chaser \chaser }(\chaser (\boldsymbol{x}+\boldsymbol{s},t))\diff^2\boldsymbol{s}& \\+ \frac{\mu_{\chaser \runner }}{\xi_{\chaser \runner }^2} \int_{\mathbb{R}^2} \boldsymbol{\underline{\underline{R}}}(\alpha_{\chaser \runner })\boldsymbol{\hat{s}}\,\Omega_{\chaser \runner }\left(\frac{s}{\xi_{\chaser \runner }}\right)g_{\chaser \runner }(\runner (\boldsymbol{x}+\boldsymbol{s},t))\diff^2\boldsymbol{s}\Biggr]\Biggr) \\
    \frac{\partial \runner }{\partial t} = D\nabla^2 \runner  -\boldsymbol{\nabla}\cdot\Biggr( \runner \changes{\phi}_\runner (\chaser ,\runner )\Biggr[\frac{\mu_{\runner \runner }}{\xi_{\runner \runner }^2} \int_{\mathbb{R}^2}\boldsymbol{\underline{\underline{R}}}(\alpha_{\runner \runner })\boldsymbol{\hat{s}}\,\Omega_{\runner \runner }\left(\frac{s}{\xi_{\runner \runner }}\right)g_{\runner \runner }(\runner (\boldsymbol{x}+\boldsymbol{s},t))\diff^2\boldsymbol{s}& \\+ \frac{\mu_{\runner \chaser }}{\xi_{\runner \chaser }^2} \int_{\mathbb{R}^2} \boldsymbol{\underline{\underline{R}}}(\alpha_{\runner \chaser })\boldsymbol{\hat{s}}\,\Omega_{\runner \chaser }\left(\frac{s}{\xi_{\runner \chaser }}\right)g_{\runner \chaser }(\chaser (\boldsymbol{x}+\boldsymbol{s},t))\diff^2\boldsymbol{s}\Biggr]\Biggr),
\end{split}
\label{eq:two_species_direct}
\end{equation}
where $D$ is the ratio of diffusivity between the two species.

In the case of gradient sensing, the governing equations are given by
\begin{equation}
\begin{split}
    \frac{\partial \chaser}{\partial t} = \nabla^2 \chaser -\boldsymbol{\nabla}\cdot\Biggr( \chaser \changes{\phi}_\chaser (\chaser ,\runner )\Biggr[\frac{\mu_{\chaser \chaser }}{\xi_{\chaser \chaser }} \,\boldsymbol{\underline{\underline{R}}}(\alpha_{\chaser \chaser })\boldsymbol{\nabla}\int_{\mathbb{R}^2}  \changes{\tilde{\Omega}}_{\chaser \chaser }\left(\frac{s}{\xi_{\chaser \chaser }}\right)g_{\chaser \chaser }(\chaser (\boldsymbol{x}+\boldsymbol{s},t))\diff^2\boldsymbol{s}& \\+ \frac{\mu_{\chaser \runner }}{\xi_{\chaser \runner }} \,\boldsymbol{\underline{\underline{R}}}(\alpha_{\chaser \runner })\boldsymbol{\nabla}\int_{\mathbb{R}^2} \changes{\tilde{\Omega}}_{\chaser \runner }\left(\frac{s}{\xi_{\chaser \runner }}\right)g_{\chaser \runner }(\runner (\boldsymbol{x}+\boldsymbol{s},t))\diff^2\boldsymbol{s}\Biggr]\Biggr) \\
    \frac{\partial \runner }{\partial t} = D\nabla^2 \runner  -\boldsymbol{\nabla}\cdot\Biggr( \runner \changes{\phi}_\runner (\chaser ,\runner )\Biggr[\frac{\mu_{\runner \runner }}{\xi_{\runner \runner }} \,\boldsymbol{\underline{\underline{R}}}(\alpha_{\runner \runner })\boldsymbol{\nabla}\int_{\mathbb{R}^2}\changes{\tilde{\Omega}}_{\runner \runner }\left(\frac{s}{\xi_{\runner \runner }}\right)g_{\runner \runner }(\runner (\boldsymbol{x}+\boldsymbol{s},t))\diff^2\boldsymbol{s}& \\+ \frac{\mu_{\runner \chaser }}{\xi_{\runner \chaser }}\,\boldsymbol{\underline{\underline{R}}}(\alpha_{\runner \chaser })\boldsymbol{\nabla} \int_{\mathbb{R}^2} \changes{\tilde{\Omega}}_{\runner \chaser }\left(\frac{s}{\xi_{\runner \chaser }}\right)g_{\runner \chaser }(\chaser (\boldsymbol{x}+\boldsymbol{s},t))\diff^2\boldsymbol{s}\Biggr]\Biggr).
\end{split}
\label{eq:two_species_gradient}
\end{equation}

For both types of models, chase-and-run dynamics are characterised by $\mu_{\chaser \runner}>0$ and $\mu_{\runner \chaser}<0$. \textbf{Fig.\, \ref{fig:chiral_run_chase_diagram}} illustrates an example of the type of dynamics that these models aim to capture.

\section{Linear analysis of pattern formation}
\label{sec:linear_analysis_pattern_formation}

\subsection{Only the parallel component of advection affects linear stability}

To find the conditions for pattern formation in these models, we perform a linear stability analysis, starting with the single species direct sensing model defined by Equation\,\eqref{eq:single_species_direct} and including the chiral rotation matrix defined by Equation\,\eqref{eq:rotation_matrix_direct}. Our approach follows the method presented in Section 3 of \citet{jewell2023}. Firstly, we linearise about the homogeneous steady-state, $\Runner$, which is determined from the total mass of the system, dictated by the initial conditions. We take $\runner(\boldsymbol{x},t)=\Runner+\tilde{\runner}(\boldsymbol{x},t)$\changes{, where $|\tilde{\runner}(\boldsymbol{x},t)|\ll 1$ is a small heterogeneous perturbation. Retaining only leading order terms yields}
\begin{equation}
    \frac{\partial \tilde{\runner}}{\partial t} = \nabla^2 \tilde{\runner} -\Runner \changes{\phi}(\Runner)\frac{\partial g}{\partial \runner}\Big\rvert_{\Runner}\,\frac{\mu}{\xi^2}\boldsymbol{\nabla}\cdot\left(\boldsymbol{\underline{\underline{R}}}(\alpha) \int_{\mathbb{R}^2} \boldsymbol{\hat{s}}\,\Omega\left(\frac{s}{\xi}\right)\tilde{\runner}(\boldsymbol{x}+\boldsymbol{s},t)\diff^2\boldsymbol{s}\right).
\end{equation}
Secondly, we expand into independent linear modes of the form $\tilde{\runner}= \runner_0 e^{i\boldsymbol{k}\cdot\boldsymbol{x}}e^{\lambda t}$, with growth rate $\lambda$ and wave-vector $\boldsymbol{k}\equiv k\boldsymbol{\hat{k}}$, with $\boldsymbol{\hat{k}}$ a unit vector. On an infinite domain, the wavenumber, $k$, is unrestricted. Consideration of a periodic finite domain simply imposes a discrete restriction on the wavenumber, so that natural numbers of the wavelength span the domain. In both cases, the dispersion relation is given by
\begin{equation}
\begin{split}
    \lambda(\boldsymbol{k}) &= -k^2 -\Runner \changes{\phi}(\Runner)\frac{\partial g}{\partial \runner}\Big\rvert_{\Runner}\,\frac{\mu}{\xi^2}\,ik\boldsymbol{\hat{k}}\cdot \boldsymbol{\underline{\underline{R}}}(\alpha)\int_{\mathbb{R}^2} \boldsymbol{\hat{s}}\,\Omega\left(\frac{s}{\xi}\right)\, e^{i\boldsymbol{k}\cdot\boldsymbol{s}} \diff^2 \boldsymbol{s}
    \\ &= -k^2 + 2\pi\,\Runner \changes{\phi}(\Runner)\frac{\partial g}{\partial \runner}\Big\rvert_{\Runner}\,\frac{\mu}{\xi^2}\,k\,(\boldsymbol{\hat{k}}\cdot \boldsymbol{\underline{\underline{R}}}(\alpha) \boldsymbol{\hat{k}})\int_0^\infty \Omega\left(\frac{s}{\xi}\right)\, s\,J_1(ks) \diff s,
\end{split}
\label{eq:dispersion_single_species_direct}
\end{equation}
\changes{where in the second line we have evaluated the angular integral, leaving only the radial integral over $s$ from the original 2D integral. In general, the $2$D Fourier transform of a function $\boldsymbol{\hat{s}}\,\Omega(s)$ is directed along $\boldsymbol{\hat{k}}$ and has magnitude proportional to the 1$^\text{st}$ order Hankel transform of $\Omega(s)$. Here, $J_1$ denotes the 1$^\text{st}$ order Bessel function of the first kind. Further details of this calculation can be found in Appendix \ref{sec:appendix_hankel}, as well as Section 3 of \citet{jewell2023}.}

Crucially, Equation \eqref{eq:dispersion_single_species_direct} differs from the dispersion relation of the non-chiral model only through the inclusion of the factor $\boldsymbol{\hat{k}}\cdot \boldsymbol{\underline{\underline{R}}}(\alpha) \boldsymbol{\hat{k}}$. This simple result holds for any constant matrix $\boldsymbol{\underline{\underline{R}}}$, making it useful for the linear analysis of models with various forms of spatial anisotropy. For instance, in certain biological systems, movement may be more favourable in the $y$ direction than in the $x$ direction, which could be encoded by a non-identity diagonal matrix. In the case of chirality, $\boldsymbol{\underline{\underline{R}}}$ corresponds to a rotation matrix, and thus
\begin{equation}
    \boldsymbol{\hat{k}}\cdot \boldsymbol{\underline{\underline{R}}}(\alpha) \boldsymbol{\hat{k}}\equiv \begin{pmatrix}
        \hat{k}_x \\ \hat{k}_y
    \end{pmatrix}^T\begin{pmatrix} \text{cos}\,\alpha & -\text{sin}\,\alpha\\\text{sin}\,\alpha & \text{cos}\,\alpha\end{pmatrix}\begin{pmatrix}
        \hat{k}_x \\ \hat{k}_y
    \end{pmatrix} = \text{cos}(\alpha).
\label{eq:kRk=cos}
\end{equation}
This indicates that chirality affects linear stability solely through the cosine of the advection angle. This result similarly applies to the gradient sensing model, where the linearised equation is given by
\begin{equation}
    \frac{\partial \tilde{\runner}}{\partial t} = \nabla^2 \tilde{\runner} -\Runner \changes{\phi}(\Runner)\frac{\partial g}{\partial \runner}\Big\rvert_{\Runner}\,\frac{\mu}{\xi}\boldsymbol{\nabla}\cdot\left(\boldsymbol{\underline{\underline{R}}}(\alpha)\boldsymbol{\nabla} \int_{\mathbb{R}^2}\changes{\tilde{\Omega}}\left(\frac{s}{\xi}\right)\tilde{\runner}(\boldsymbol{x}+\boldsymbol{s},t)\diff^2\boldsymbol{s}\right).
\end{equation}
Expanding into linear modes yields
\begin{equation}
    \lambda(\boldsymbol{k}) = -k^2 + 2\pi\,\Runner \changes{\phi}(\Runner)\frac{\partial g}{\partial \runner}\Big\rvert_{\Runner}\,\frac{\mu}{\xi}\,k^2\,(\boldsymbol{\hat{k}}\cdot \boldsymbol{\underline{\underline{R}}}(\alpha) \boldsymbol{\hat{k}})\int_0^\infty \changes{\tilde{\Omega}}\left(\frac{s}{\xi}\right)\, s\,J_0(ks) \diff s,
\label{eq:dispersion_single_species_gradient}
\end{equation}
where $J_0(ks)$ is a 0$^\text{th}$ order Bessel function of the first kind, and the integral transform corresponds to a 0$^\text{th}$ order Hankel transform. Once again, chirality only affects the dispersion relation through the cosine of the angle, as $\boldsymbol{\hat{k}}\cdot \boldsymbol{\underline{\underline{R}}}(\alpha) \boldsymbol{\hat{k}} = \text{cos}(\alpha)$. These results are consistent with the stability conditions found for the local model of \citet{woolley2017chiral}, whose angular terms also only appear through the cosine.

\textbf{Thus, for these chiral models, only the component of the induced advection that is parallel to the line of separation contributes to linear stability}. In Section\,\ref{sec:geometric_argument}, we provide an intuitive geometric interpretation of this result. Notably, this result also extends to two-species models.

\subsection{Two species dispersion relation}

For the two species models, we can follow a similar process, starting with either Equations \eqref{eq:two_species_direct} or \eqref{eq:two_species_gradient}, for the direct sensing or gradient sensing models, respectively. First, we linearise the governing equations about the homogeneous steady-state, which is determined by the total mass from the initial conditions. We set
\begin{equation}
    \begin{pmatrix}\chaser(\boldsymbol{x}, t) \\\runner(\boldsymbol{x}, t) \end{pmatrix} = \begin{pmatrix}\Chaser \\ \Runner \end{pmatrix}+\begin{pmatrix}\tilde{\chaser}(\boldsymbol{x}, t) \\ \tilde{\runner}(\boldsymbol{x}, t) \end{pmatrix}.
\end{equation}
Next, we expand the perturbation into linear modes of the form
\begin{equation}
    \begin{pmatrix}\tilde{\chaser}(\boldsymbol{x}, t) \\ \tilde{\runner}(\boldsymbol{x}, t) \end{pmatrix} = e^{i\boldsymbol{k}\cdot\boldsymbol{x}}e^{\lambda t} \begin{pmatrix} a_\chaser\\ a_\runner\end{pmatrix},
\end{equation}
where $\frac{a_\chaser}{a_\runner}$ encodes  the relative amplitude of the chaser density to the runner density in the mode. Substituting this into the linearised equations yields the dispersion relation
\begin{equation}
    \lambda \begin{pmatrix} a_\chaser\\ a_\runner\end{pmatrix} = \begin{pmatrix} -k^2 + \Lambda_{\chaser \chaser}(k) & \Lambda_{\chaser \runner}(k) \\\Lambda_{\runner \chaser}(k)& -Dk^2 + \Lambda_{\runner \runner}(k)\end{pmatrix}\begin{pmatrix} a_\chaser\\ a_\runner\end{pmatrix} \equiv \underline{\underline{\boldsymbol{F}}}(k)\begin{pmatrix} a_\chaser\\ a_\runner\end{pmatrix}.
\label{eq:matrix_dispersion}
\end{equation}
Here, each $\Lambda_{uv}$ is the contribution from the nonlocal interaction of species $v$ on species $u$, where $u, v \in \{\chaser,\runner\}$ with \textit{corresponding} homogeneous states $U,V \in \{\Chaser,\Runner\}$.

For the direct sensing model,
\begin{equation}
    \Lambda_{uv}(k) \equiv \text{cos}(\alpha_{uv})\,2\pi U\changes{\phi}_u(\Chaser,\Runner)\frac{\partial g_{uv}}{\partial v}\Big\rvert_{V} \frac{\mu_{uv}}{\xi_{uv}^2}\,k\int_{0}^{\infty}\Omega_{uv}\left(\frac{s}{\xi_{uv}}\right)s\,J_1(ks)\changes{\,\diff s},
\label{eq:Lambda_direct}
\end{equation}
and for the gradient sensing model,
\begin{equation}
    \Lambda_{uv}(k) \equiv \text{cos}(\alpha_{uv})\,2\pi U\changes{\phi}_u(\Chaser,\Runner)\frac{\partial g_{uv}}{\partial v}\Big\rvert_{V} \frac{\mu_{uv}}{\xi_{uv}}\,k^2\int_{0}^{\infty}\changes{\tilde{\Omega}}_{uv}\left(\frac{s}{\xi_{uv}}\right)s\,J_0(ks)\changes{\,\diff s},
\label{eq:Lambda_gradient}
\end{equation}
by analogy with Equations \eqref{eq:dispersion_single_species_direct} and \eqref{eq:dispersion_single_species_gradient}, respectively. Both models share similar dispersion relations, with the only notable difference being the order of the Hankel transform of the interaction kernel, in their definitions of $\Lambda_{uv}$. \changes{The link between the two frameworks can be seen by choosing $\Omega_{uv} = \frac{\diff}{\diff s}\tilde{\Omega}_{uv}$ and integrating Equation\,\eqref{eq:Lambda_direct} by parts, which will yield Equation\,\eqref{eq:Lambda_gradient}.} Crucially, in both equations, chirality only features through the cosine of the advection angle.

The growth rates, $\lambda(k)$, are the eigenvalues of $\underline{\underline{\boldsymbol{F}}}(k)$, and are thus given by
\begin{equation}
    \lambda_{\pm}(k) = \frac{1}{2}\left(-(1+D)k^2 + (\Lambda_{\chaser\chaser}+\Lambda_{\runner\runner}) \pm \sqrt{4\Lambda_{\chaser \runner}\Lambda_{\runner \chaser}+[(1-D)k^2+(\Lambda_{\runner \runner}-\Lambda_{\chaser \chaser})]^2}\right).
\label{eq:lambda}
\end{equation}

Equation \eqref{eq:lambda} is the dispersion relation for our two-species chiral models. It reveals whether patterns can autonomously form from small spatially heterogeneous perturbations about a spatially uniform steady-state. Patterns can form when these perturbations can grow, which requires that $\text{Re}(\lambda_+)>0$ for some $k>0$. 
Furthermore, we expect these patterns to be stationary at early time if $\text{Im}(\lambda_+)=0$, which occurs if and only if
\begin{equation}
    4\,\Lambda_{\chaser \runner}\Lambda_{\runner \chaser}+[(1-D)k^2+(\Lambda_{\runner \runner}-\Lambda_{\chaser \chaser})]^2 \geq 0.
\label{eq:oscillation_conditions}
\end{equation}
If instead this discriminant is negative, then $\text{Im}(\lambda_+)\neq 0$, and we expect patterns that initially oscillate in time. At early times, the amplitude of each mode in the perturbation is small, and so the dynamics are accurately captured by a linear approximation. Accordingly, early times are often referred to as the `linear regime'.

To gain insight from these equations, we will assume that each $\Lambda_{uv}(k)$ shares the same sign as $\mu_{uv}$ for values of $k$ that are `relevant for pattern formation' -- typically those corresponding to the largest positive values of $\lambda(k)$.  From Equations \eqref{eq:Lambda_direct} and \eqref{eq:Lambda_gradient}, we see that this assumption requires that: (a) $\changes{\phi}_u(\Chaser,\Runner)\frac{\partial g_{uv}}{\partial v}\Big\rvert_{V}>0$, which follows from our core model assumptions, and (b) the Hankel transforms of the interaction kernels are positive at these key values of $k$. \changes{We argue in Appendix \ref{sec:appendix_kernel_assumptions} that this latter requirement is satisfied in many biologically relevant cases.}

\section{Geometric interpretation of linear analysis}
\label{sec:geometric_argument}
In this section, we provide an intuitive argument, based on simple geometry, for why only the parallel component of the induced advection affects linear stability and capacity for pattern formation. We show that if a chiral interaction generates movement in a direction perpendicular to the separation of the two interacting agents, then this perpendicular movement does not drive any growth or decay of aggregations of these agents. In other words, we illustrate why chirality only enters the dispersion relations through the cosine of the interaction angle.

\subsection{Direct sensing}
\begin{figure}
    \centering
    \includegraphics[width=1\textwidth]{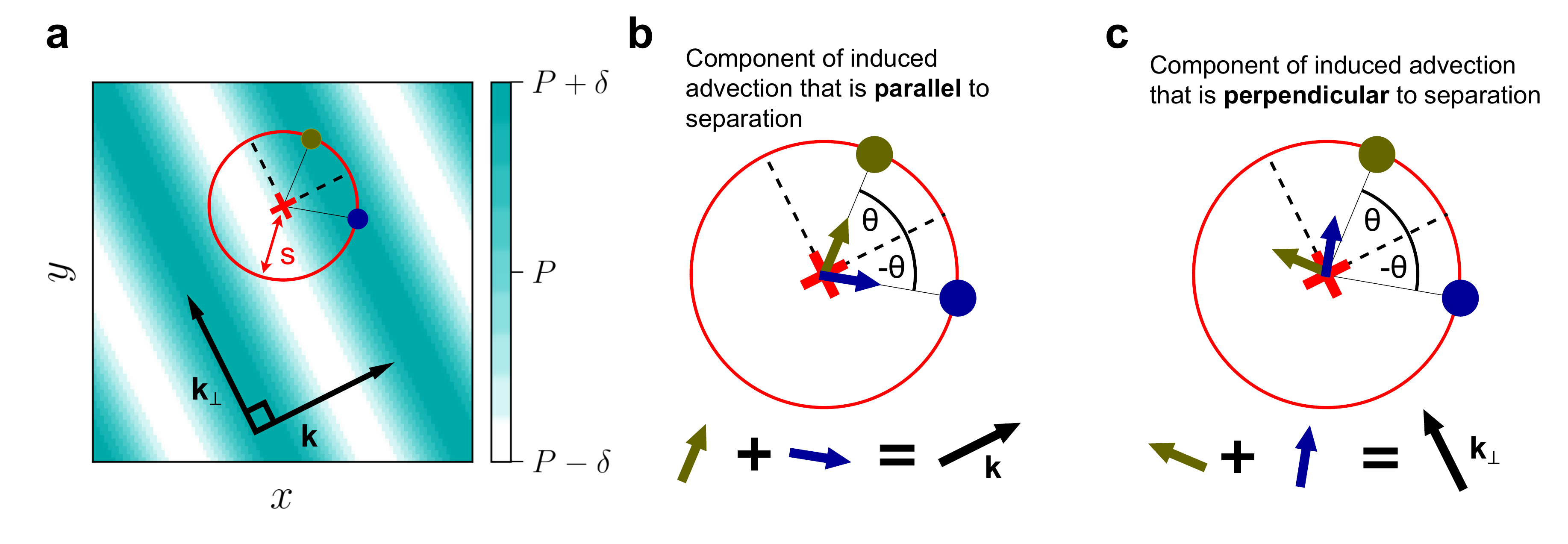}
    \caption{Schematic illustrating how the parallel component of a chiral interaction affects linear stability and pattern formation, whereas the perpendicular component does not. \textbf{a} A small sinusoidal perturbation of the density of a single-species with amplitude $\delta$ and wave-vector $\boldsymbol{k}$. We consider the advection at an arbitrary point (red `$\times$') induced by interactions with points at distance $s$ (red circle). Points (dark yellow and blue) with opposite angles to the $\boldsymbol{k}$ direction are paired up, as they have the same density and so induce the same magnitude of advection. \textbf{b} Parallel interactions create a net advection parallel to the $\boldsymbol{k}$ direction, thereby changing the density distribution, affecting linear stability. \textbf{c} Perpendicular interactions create a net advection perpendicular to the $\boldsymbol{k}$ direction, and so only cause movement in directions that do not change the density distribution, and so do not affect linear stability.}
    \label{fig:geometric_interpretation}
\end{figure}

Consider a single species distributed homogeneously in space with density $P$. Then, as shown in \textbf{Fig.\,\ref{fig:geometric_interpretation}\,a}, consider a small perturbation in this density with amplitude $\delta$ that varies in space sinusoidally with wave-vector $\boldsymbol{k}$. If the perturbation grows, there is pattern formation. If it decays, the state returns to being spatially homogeneous. We can determine which outcome will occur by considering the advection induced at an arbitrary point, such as the red `$\times$' in \textbf{Fig.\,\ref{fig:geometric_interpretation}}. Advection at `$\times$' is induced by nonlocal interactions with agents at surrounding points. Consider points that are at some arbitrary distance, $s$, from `$\times$'. These are represented by the red circle in \textbf{Fig.\,\ref{fig:geometric_interpretation}}.

Now consider the dark yellow and blue points in \textbf{Fig.\,\ref{fig:geometric_interpretation}}, which are at a distance $s$ and are at an arbitrary angle $\theta$ and $-\theta$ relative to the direction of heterogeneity of the perturbation, $\boldsymbol{k}$. The advection induced at `$\times$' by each point is shown as the arrow with corresponding colour in \textbf{Fig.\,\ref{fig:geometric_interpretation}\,b} and \textbf{c}. The advection induced by both points will have the same magnitude, because there is an equal density of agents at both points. Thus, when advection is induced parallel to separation, we see that the overall net advection from both points is directed parallel to $\boldsymbol{k}$. In contrast, when advection is induced perpendicular to separation, the overall net advection from both points is directed perpendicular to $\boldsymbol{k}$. The angle $\theta$ is arbitrary, and so all points at distance $s$ (on the red circle) can be paired up in this way, such that the two points induce the same magnitude of advection and their net advection is parallel to $\boldsymbol{k}$ in the parallel case and perpendicular to $\boldsymbol{k}$ in the perpendicular case. The red `$\times$' point and the distance `s' are also arbitrary, so this argument applies to all points interacting at all distances.

Overall, this means that the component of induced advection that is perpendicular to separation only drives movement along directions where the density of agents does not vary, i.e. `up or down the stripe'. It therefore has no affect on the density distribution of agents, and thus cannot affect linear stability or pattern formation. In contrast, the component of induced advection that is parallel to separation only drives movement in the direction of heterogeneity in a perturbation. Agents will therefore move to regions with differing densities, leading to either growth or decay in the perturbation, and thus affecting linear stability and pattern formation.

\subsection{Gradient sensing and assumptions}
The argument presented above refers to the direct sensing model. Nevertheless, the conclusions also hold for the gradient sensing model. This can be shown directly, or by simply noting the relationship between the gradient and direct sensing models. For the former, consider the region within a circle around an arbitrary point, such as the red circle around the red `$\times$' in \textbf{Fig.\,\ref{fig:geometric_interpretation}}. In the gradient sensing model, advection depends on the density gradient averaged across such a region. By symmetry, density only varies in the $\boldsymbol{k}$ direction, and so the gradient must point in this direction. Similarly to in the direct sensing model, advection that is induced parallel to this gradient is thus directed parallel to $\boldsymbol{k}$, affecting linear stability. Once again, advection induced perpendicular to the gradient is directed perpendicular to $\boldsymbol{k}$, and does not affect linear stability.

The above arguments rely on having a density perturbation that only varies in one direction. This is justified because linear modes are orthogonal for arbitrarily small perturbations, allowing us to treat each mode independently, and each mode indeed only varies in one direction.

\subsection{Further insights}
Beyond linearity, these arguments also tell us about the role of chiral interactions even in large amplitude patterns. A classical stripe pattern indeed only varies in one direction. We therefore can deduce that the parallel component of interaction is able to hold the stripe together. In contrast, the perpendicular component of interaction induces movement of agents along the stripe (assuming there is no force to oppose this movement, such as volume-filling).  This implies that the equilibrium can be dynamic: the density of agents along the stripe is unchanging, but individual agents may be travelling along the stripe. The same arguments can be adapted to a spot pattern, implying that agents travel around the spot at a constant radius, without changing the density distribution.

This also suggests that the choice of boundary conditions could significantly alter the available patterns. With a periodic boundary, a stripe can maintain dynamic equilibrium as there are always agents available from one end of the stripe and there is always space for them to move to at the other end.  With no flux boundary conditions, this is no longer true. This may suggest that, in systems with chiral interactions, simple stripe patterns become less viable on a closed finite domain.

\section{Insights on pattern formation}
\label{sec:insights_PF}

\begin{figure}
    \centering
    \includegraphics[width=0.7\linewidth]{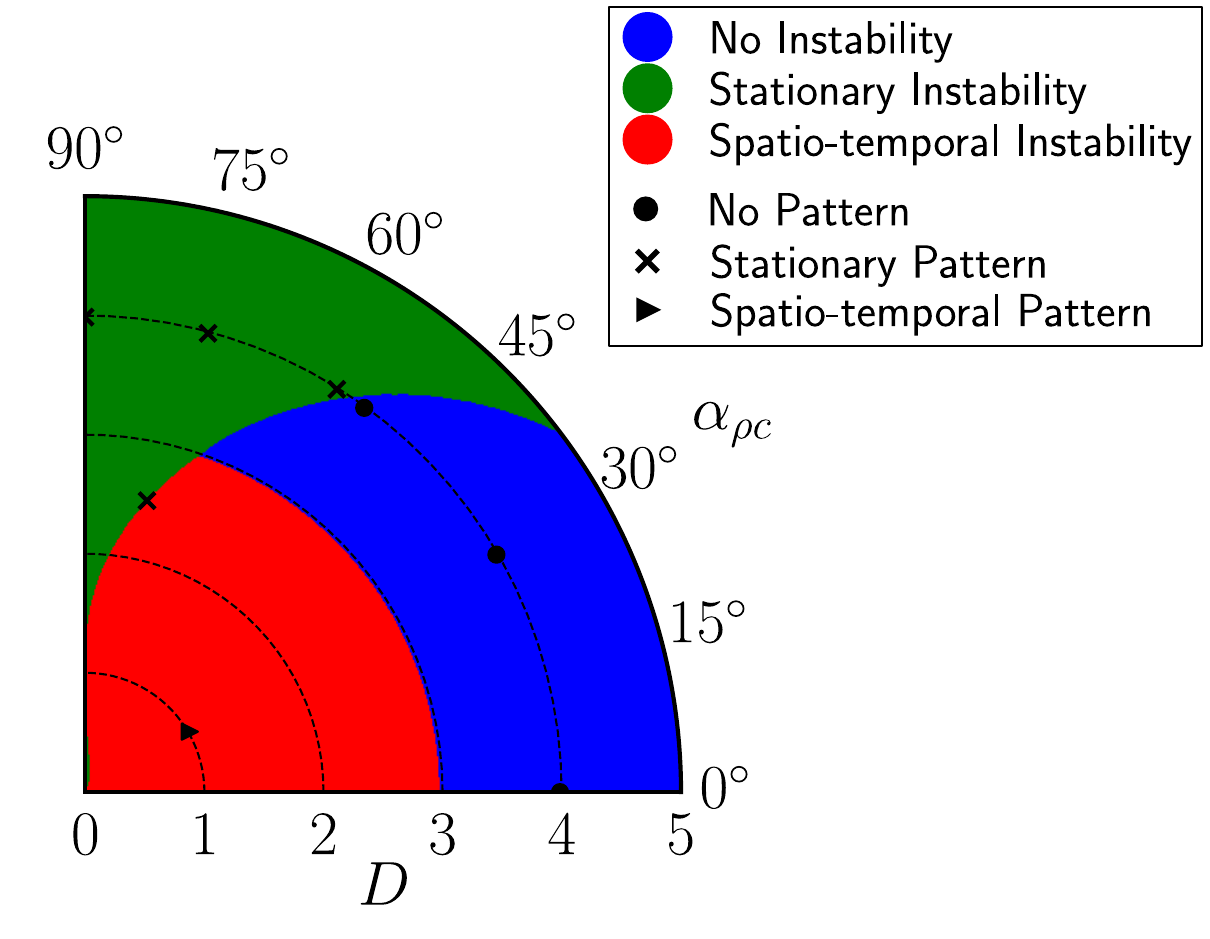}
    \caption{\changes{Classification of the ($D$, $\alpha_{\runner \chaser}$) parameter space in polar coordinates, based on the predictions of Equation \eqref{eq:lambda}, on a domain of size $L=8$. Increasing the magnitude of the chiral running angle, $\alpha_{\runner \chaser}$, can shift the system from a linearly stable homogeneous state ($\text{Re}[\lambda(k)]<0$, blue) or a spatio-temporal instability ($\text{Re}[\lambda(k)]>0$, $\text{Im}[\lambda(k)]\neq 0$, red) to a stationary instability ($\text{Re}[\lambda(k)]>0$, $\text{Im}[\lambda(k)]=0$, green). Black $\circ, \times, \blacktriangleright$ markers indicate parameter values tested in numerical simulations and their patterning outcome. For this figure, $\alpha_{\runner \runner}=\alpha_{\chaser \chaser}=\alpha_{\chaser \runner}=0$, with all other parameters given in Table \ref{tab:parameters} of Appendix \ref{sec:appendix_parameter_values}.}}
    \label{fig:phase_map}
\end{figure}

\changes{Before presenting numerical simulations, it is useful to first highlight the key conceptual insights from the linear stability analysis. This provides a framework for understanding how chirality and chase-and-run dynamics influence pattern formation and sets the stage for interpreting the subsequent full nonlinear dynamics. In particular,} the linear stability analysis in Section\,\ref{sec:linear_analysis_pattern_formation} tells us that pattern formation occurs if $\text{Re}(\lambda_+)>0$ for some $k>0$, where $\lambda_+(k)$ is specified by Equation \eqref{eq:lambda}. Additionally, we expect patterns to be stationary (rather than oscillating in time) if Condition \eqref{eq:oscillation_conditions} is satisfied, that is, if $\text{Im}(\lambda_+(k))=0$.

We know that each nonlocal interaction only affects $\lambda$ via $\text{cos}(\alpha_{uv})$, i.e. through its parallel component of induced advection. \textbf{Therefore, the impact of any interaction on the capacity for pattern formation, and on whether the pattern is stationary or oscillatory, can be suppressed by making the interaction more perpendicular.} Thus, chirality can promote pattern formation by inhibiting pattern-suppressing interactions. In the extreme case, any interaction that induces purely perpendicular movement ($\alpha_{uv}=90^{\circ}$) will have no effect on linear stability.

Chase-and-run interactions are defined by $\mu_{\chaser \runner}>0$ and $\mu_{\runner \chaser}<0$, which means that $\Lambda_{\chaser \runner}\Lambda_{\runner \chaser}<0$. From Equation \eqref{eq:lambda}, these interactions therefore can only decrease $\text{Re}(\lambda_+)$, thereby suppressing pattern formation. Additionally, Condition\,\eqref{eq:oscillation_conditions} implies that these interactions promote temporal oscillations in patterns which are able to form. In the context of developmental biology, both effects are typically detrimental to the creation of stationary structures within an organism. 

Chirality presents a possible mechanism for suppressing these `detrimental' effects of chase-and-run dynamics. If the runner or chaser species were to move more chirally, increasing the angle $\alpha_{\runner \chaser}$ or $\alpha_{\chaser \runner}$, then the magnitude of the $\Lambda_{\chaser \runner}\Lambda_{\runner \chaser}$ term would be reduced, which both promotes pattern formation and suppresses oscillations. \changes{This effect is illustrated in the phase diagram in \textbf{Fig.\,\ref{fig:phase_map}}, where increasing $\alpha_{\runner \chaser}$ shifts the system out of regimes with either a stable homogeneous state or a spatio-temporal instability and into a regime with a stationary (non-temporal) instability.}

However, while chirality suppresses one mechanism of temporal oscillations --- namely, those arising through linear interactions --- it does not eliminate all possible sources of oscillations. When the amplitude of the pattern is sufficiently large, nonlinear effects can sometimes introduce oscillations. The converse is also true: nonlinear effects can sometimes destroy oscillations, leading to a stationary steady-state. Thus, chirality can be said to promote, but not necessarily guarantee, stationary pattern formation.

Another limitation is that chirality cannot reverse the effects of an interaction -- it can only weaken them. For example, chirality can mitigate the pattern-suppressing influence of chase-and-run interactions, but it cannot cause such interactions to actively promote patterning. An additional mechanism, such as self-species interactions, is still necessary to drive the instability required for pattern formation.

Interestingly, a mechanism which acts in direct opposition to the effects of chase-and-run interactions is differential self-interaction. Differential self-interaction, where self-interactions between chasers differ in magnitude and/or direction to self-interactions between runners, is encoded in the term $[(1-D)k^2+(\Lambda_{\runner \runner}-\Lambda_{\chaser \chaser})]^2$ in Equations \eqref{eq:lambda} and \eqref{eq:oscillation_conditions} (note that we include diffusion in this consideration of self-interaction). It is strictly positive, whereas chase-and-run terms, $\Lambda_{\chaser \runner}\Lambda_{\runner \chaser}$, are strictly negative. Differential self-interaction thus promotes pattern formation where chase-and-run suppresses it. Furthermore, the balance of these two effects determines whether emergent patterns are stationary or oscillatory (in the linear regime), as given by Condition\,\eqref{eq:oscillation_conditions}. This condition depends only on the magnitude of the differential self-interaction, and not on whether it is the chaser or the runner that has the stronger interaction.

An advantage of our linear analysis is that its insights are fairly general. It applies to both the direct sensing model and the gradient sensing model. It also requires minimal assumptions about the details of the volume-filling, interaction kernels, and $g_{uv}$ functions. However, recent work by \citet{krause2024turing_conditions_not_enough} and \citet{oliver2024_turing_arent_enough} has shown that predictions from linear analysis do not always extend to the nonlinear regime, where pattern amplitudes are large enough for nonlinear effects to become significant. In some examples from these studies, patterns predicted by linear theory are only transient, with the dynamics eventually settling on a different homogeneous steady-state. In our case, however, such behaviour is unlikely to occur, as Equations \eqref{eq:two_species_direct} and \eqref{eq:two_species_gradient} have only one viable homogeneous state, dictated by mass conservation. Nevertheless, in Section \ref{sec:simulations_1}, we test our insights from linear analysis using numerical simulations of the full nonlinear systems. 

\begin{figure}
    \centering
    \includegraphics[width=1\textwidth]{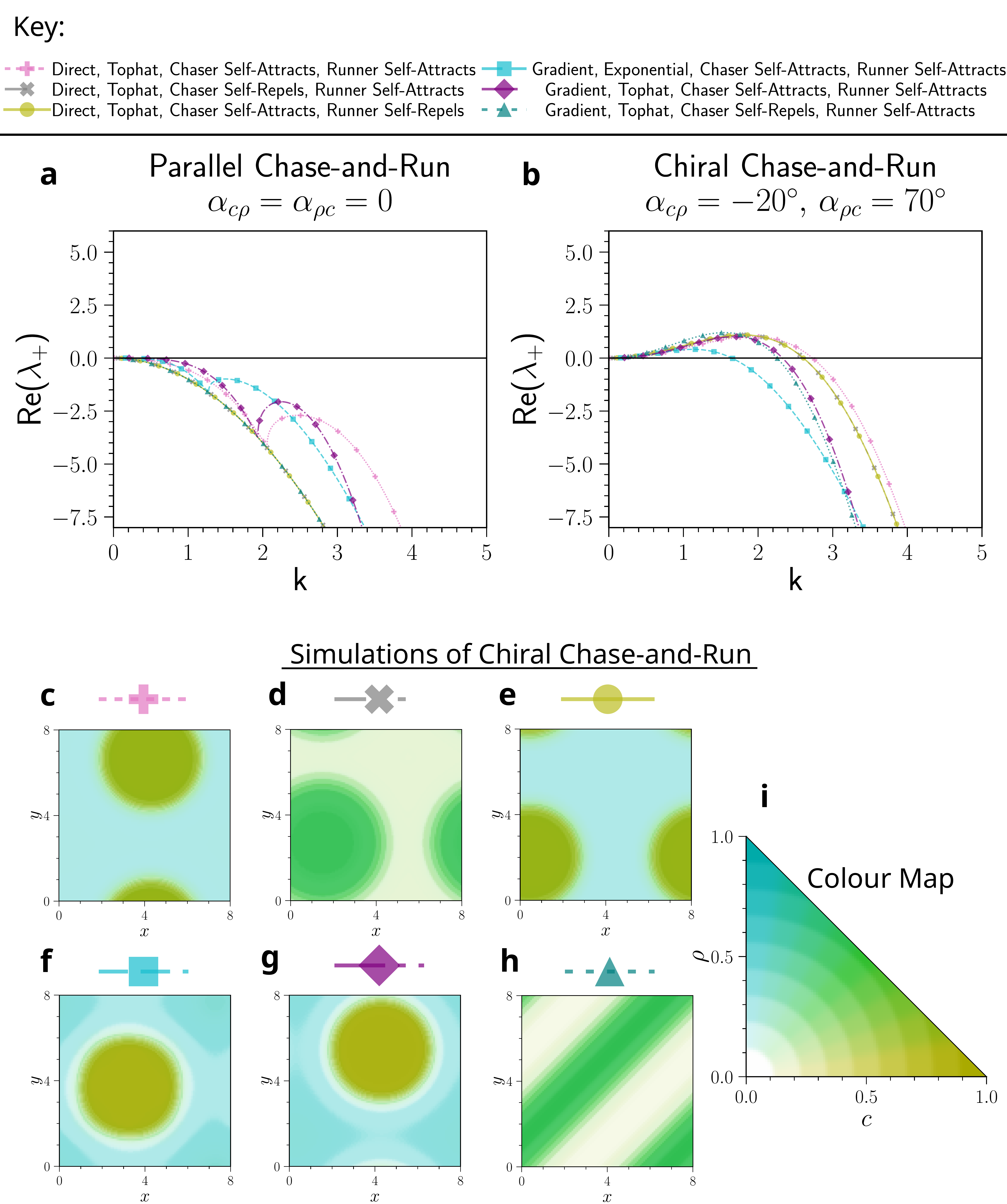}
    \caption{Examples in which patterns fail to form with parallel chase-and-run dynamics but emerge when chase-and-run is chiral ($\alpha_{\chaser\runner}=-20^{\circ}$ and $\alpha_{\runner\chaser}=70^{\circ}$). This behaviour occurs across various model sensing types, kernels, and parameters (see Key). Dispersion curves show that the homogeneous steady-state is linearly stable for parallel chase-and-run (\textbf{a}) but unstable for chiral chase-and-run (\textbf{b}). Accordingly, simulations of cases in \textbf{a} remained homogeneous (not shown), while those of \textbf{b} formed patterns. Heatmaps of these patterns at late times are shown in \textbf{c}-\textbf{h}, with simulations run to $t=500$. Parameter values are in Table \ref{tab:parameters} of Appendix \ref{sec:appendix_parameter_values}. Heatmaps use the 2D colourmap in \textbf{i}, where high chaser density is yellow, high runner density is blue, and both species overlapping is green.}
    \label{fig:chirality_promotes_pf}
\end{figure}

\begin{figure}
    \centering
    \includegraphics[width=0.95\textwidth]{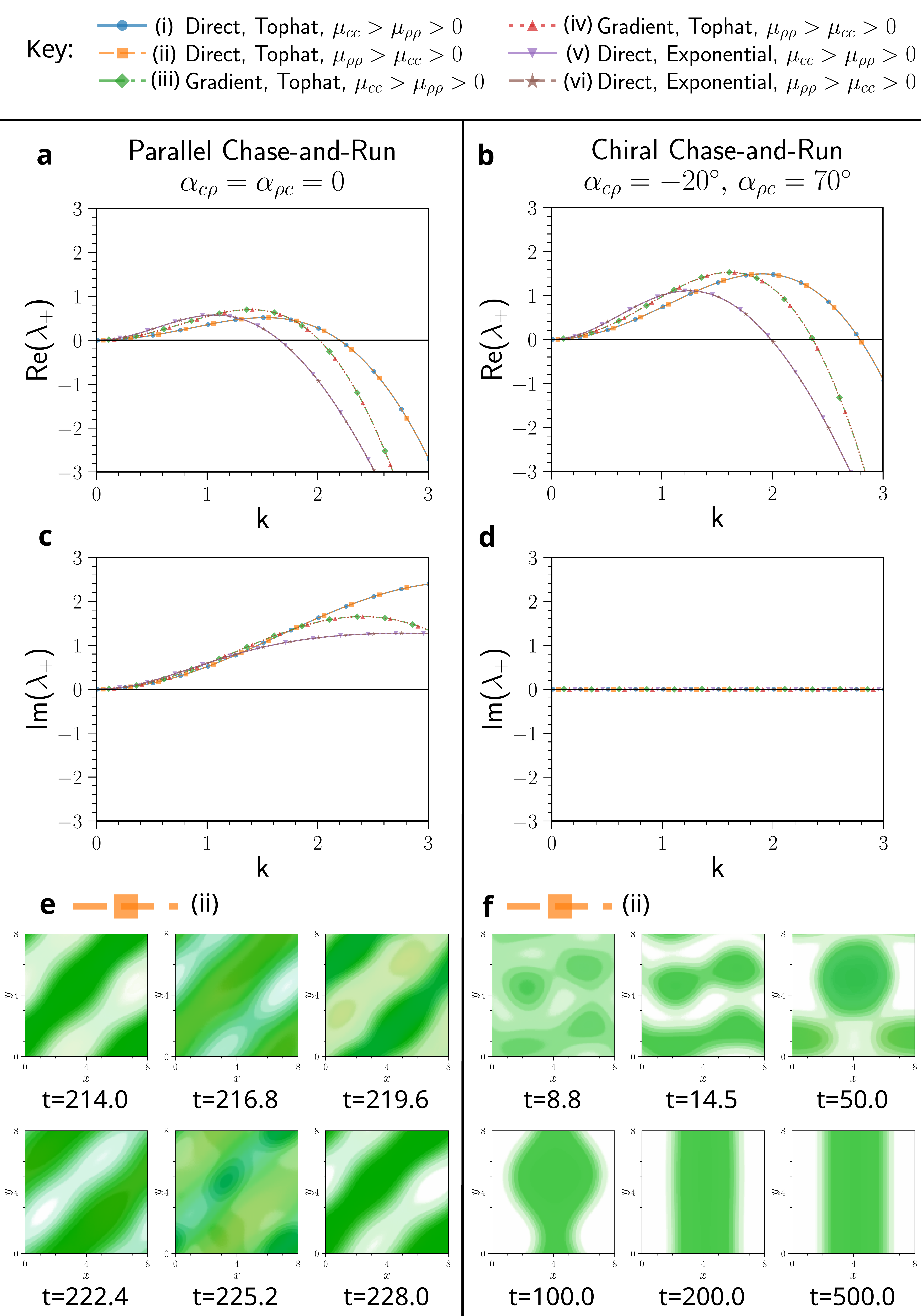}
    \caption{Examples in which patterns oscillate in time with parallel chase-and-run dynamics but are stationary when chase-and-run is chiral ($\alpha_{\chaser\runner}=-20^{\circ}$ and $\alpha_{\runner\chaser}=70^{\circ}$). This behaviour occurs across various model sensing types, kernels, and parameters (see Key). Dispersion curves show that the homogeneous steady-state is linearly unstable in all cases (\textbf{a}, \textbf{b}), but the imaginary part of $\lambda$ is non-zero for parallel chase-and-run (\textbf{c}) and zero for chiral chase-and-run (\textbf{d}). An example simulation of parallel chase-and-run (\textbf{e}) exhibits persistent oscillations, whereas the same system with chirality (\textbf{f}) reaches a stationary state that is stable for the remainder of the simulation. \textbf{e} and \textbf{f} use the colourmap from \textbf{Fig.\,\ref{fig:chirality_promotes_pf}}\,\textbf{i}, but in \textbf{e}, colours are rescaled for visibility ($\chaser_\text{min},\, \runner_\text{min}=0.05$ and $\chaser_\text{max},\, \runner_\text{max}=0.54$). Parameter values for each case (i–vi) are in Table \ref{tab:parameters} of Appendix \ref{sec:appendix_parameter_values}. Videos of these simulations are available at \citet{My_GitHub}.}
    \label{fig:chirality_suppresses_oscillations}
\end{figure}

\begin{figure}
    \centering
    \includegraphics[width=1\textwidth]{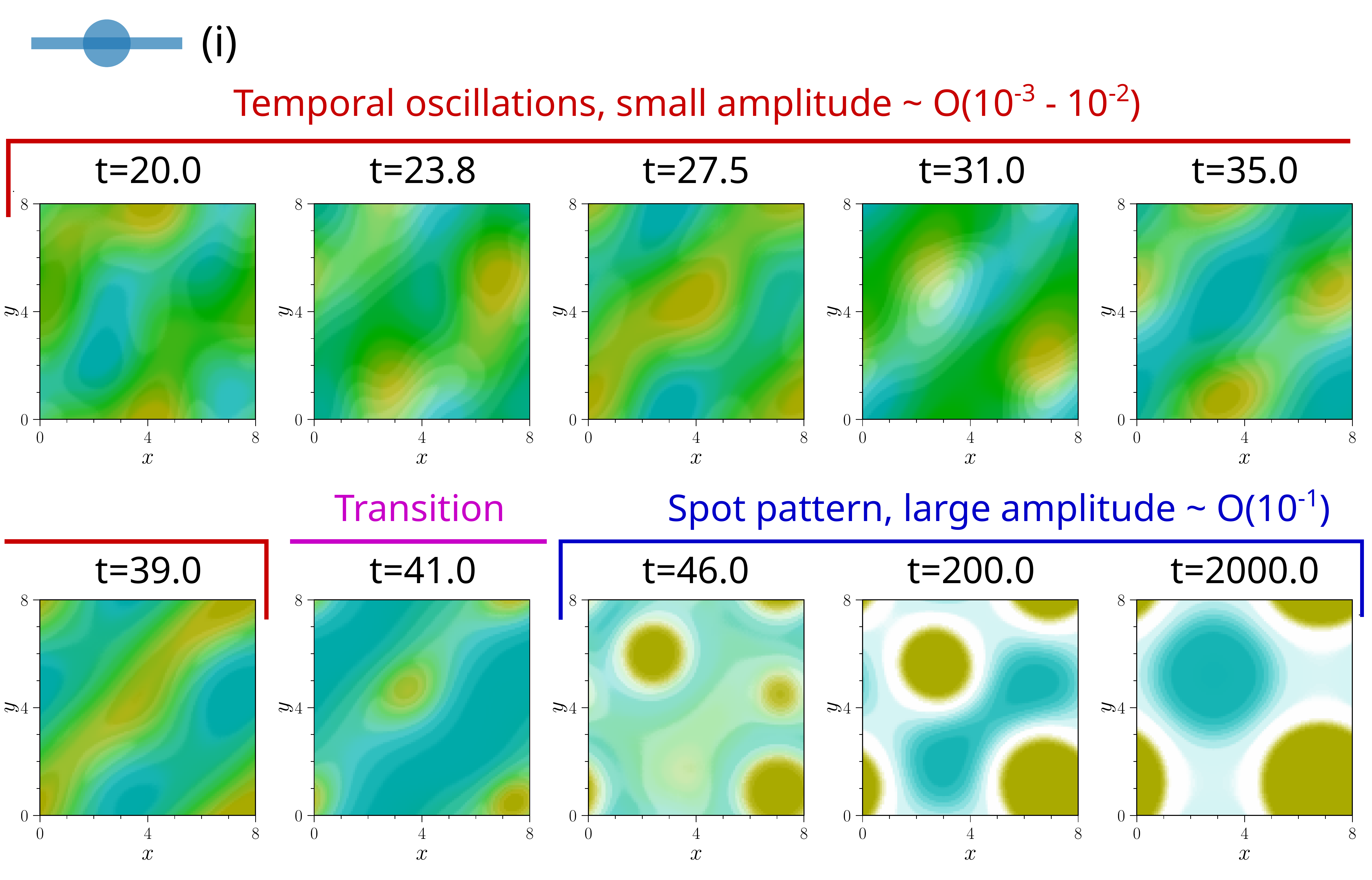}
    \caption{Simulation of parallel chase-and-run dynamics, initially exhibiting temporal oscillations before transitioning into a spot pattern. The spots then merge into a single larger spot for each species, reaching a state that is stationary at late times. In this case, linear theory predicts a complex growth rate ($\text{Im}(\lambda)\neq 0$), explaining the initial oscillation but failing to capture the long time behaviour. During the oscillatory and transition stages ($t=0$ to $t=41$), pattern amplitudes are small, $\mathcal{O}(10^{-3})$ to $\mathcal{O}(10^{-2})$, so colours from the colourmap in \textbf{Fig.\,\ref{fig:chirality_promotes_pf}}\,\textbf{i} are rescaled for visibility. After the pattern collapses into spots ($t=46$), amplitudes become large, $\mathcal{O}(10^{-1})$, and the standard $[0,1]^2$ colourmap range is used. Parameters match those of \textbf{Fig.}\,\ref{fig:chirality_suppresses_oscillations} case (i), and are listed in Table \ref{tab:parameters} of Appendix \ref{sec:appendix_parameter_values}. A Video of this simulation is available at \citet{My_GitHub}.}
    \label{fig:oscillations_become_stationary}
\end{figure}

\section{Simulations of pattern formation}
\label{sec:simulations_1}
\changes{To test the predictions from our linear analysis, we simulate the full nonlinear integro-PDEs, namely Equation \eqref{eq:two_species_direct} for the direct sensing model, and Equation \eqref{eq:two_species_gradient} for the gradient sensing model. These simulations allow us to assess how well the conceptual insights from the linear analysis extend into the fully nonlinear regime and to observe phenomena that arise beyond the scope of linear theory.}

\subsection{Choice of nonlinear functions}
\changes{For all simulations in this study, we} specify the volume-filling functions as $\changes{\phi}_\chaser(\chaser, \runner)=\changes{\phi}_\runner(\chaser, \runner)=1-\chaser-\runner$, and the $g_{uv}$ functions as $g_{\chaser \chaser}(\chaser)=g_{\runner \chaser}(\chaser)=\chaser$,\, $g_{\chaser \runner}(\runner)=g_{\runner \runner}(\runner)=\runner$. \changes{For the interaction kernels, we consider two commonly studied functions: the exponential kernel, 
\begin{equation}
    \Omega\left(\frac{s}{\xi}\right) = \frac{1}{2\pi} e^{-\frac{s}{\xi}},
\label{eq:exponential_kernel}
\end{equation}
and the tophat kernel,
\begin{equation}
    \Omega\left(\frac{s}{\xi}\right)= \begin{cases}
        \frac{1}{\pi},\quad s\leq \xi \\ 0,\quad s>\xi \,.
    \end{cases}
\label{eq:top_hat_kernel}
\end{equation}
\changes{In different examples, the interaction kernel is taken as either tophat or exponential, for both the direct sensing model ($\Omega$) and the gradient sensing model ($\tilde{\Omega}$).} For simulations using the exponential kernel, the kernel's spatial extent is cut-off at half the domain length ($\frac{L}{2}$) in all directions. This ensures that any two points only interact with each other once, rather than wrapping round the domain to interact from multiple directions. We have confirmed that this choice has a negligible impact on the linear stability analysis for all the examples in this study, as the domain length $L$ is always suitably larger than the decay scale of the exponential $\xi$.}

\subsection{Numerical scheme}
As a proxy for the infinite domain considered in our linear analysis, we choose periodic boundary conditions on a square domain, $[0,L]^2$. The initial conditions are taken as a homogeneous state $(\Chaser, \Runner)$, where $C$ and $P$ are typically of order $\mathcal{O}(10^{-1})$, with a perturbation of random Gaussian noise with zero mean and standard deviation $10^{-3}$. For the numerical integration, we first discretise space into a grid of $100\times 100$ mesh points, and then integrate the $100^2$ ODEs in time using \texttt{SciPy}'s BDF1-5 implicit variable-order scheme. The nonlocal term is computed using a fast Fourier transform convolution method. See \citet{My_GitHub} for further details of the numerical scheme.

\subsection{Results}

\textbf{Fig.\,\ref{fig:chirality_promotes_pf}} shows examples of systems in which parallel chase-and-run dynamics prevent pattern formation, while the same parameters with chiral chase-and-run lead to pattern formation. This is predicted by the dispersion relations (\textbf{a}, \textbf{b}) and is confirmed by simulations, where patterns only emerge under chiral dynamics ($\textbf{c-h}$). For the chiral simulations, the chasing and running angles are set to $\alpha_{\chaser\runner}=-20^{\circ}$ and $\alpha_{\runner\chaser}=70^{\circ}$, although any sufficiently large angles would enable pattern formation. As evidenced in \textbf{Figs.\,\ref{fig:chirality_promotes_pf}} \changes{and \textbf{\ref{fig:phase_map}}}, this is a general phenomenon holding across different model types (direct and gradient sensing), kernel functions (tophat and exponential), as well as different combinations of self-attraction and self-repulsion in the chasers and runners. These results support our linear analysis prediction that chirality promotes pattern formation in chase-and-run systems.

We also provide evidence for the second key prediction: chirality can suppress temporal oscillations. \textbf{Fig.\,\ref{fig:chirality_suppresses_oscillations}} shows cases where parallel chase-and-run dynamics yield dispersion relations with $\text{Re}(\lambda_+)>0$ (\textbf{a}) and $\text{Im}(\lambda_+)\neq 0$ (\textbf{c}), suggesting patterns that oscillate in time. Using chiral chase-and-run, with all other parameters fixed, results in $\text{Re}(\lambda_+)>0$ (\textbf{b}) but $\text{Im}(\lambda_+)= 0$ (\textbf{d}), predicting stationary patterns. 

Simulations of parallel chase-and-run in these cases all exhibited initial oscillations, with structures forming, breaking apart or translating across the domain. Some, like \textbf{Fig.\,\ref{fig:chirality_suppresses_oscillations}\,e}, showed persistent oscillations. Others, such as \textbf{Fig.\,\ref{fig:oscillations_become_stationary}}, displayed transient oscillations that eventually collapsed into stationary spot or stripe patterns. In contrast, the chiral systems in \textbf{Fig.\,\ref{fig:chirality_suppresses_oscillations}} showed no oscillations. Instead, as seen in \textbf{Fig.\,\ref{fig:chirality_suppresses_oscillations}\,f}, structures formed, remained intact with minimal movement, and then, only after reaching a large amplitude ($\chaser, \runner \sim\mathcal{O}(10^{-1})$), coalesced into a single spot or stripe that remained stable seemingly indefinitely.

Overall, these simulations confirm that chirality both enables pattern formation and suppresses one mechanism of temporal oscillations, as predicted by linear theory. However, cases where oscillations predicted by linear analysis disappeared at large pattern amplitudes, such as in \textbf{Fig.\,\ref{fig:oscillations_become_stationary}}, highlight how such linear analysis can fail to capture long time behaviour.

Another notable result from these simulations is that every stationary late-time pattern in \textbf{Figs.}\,\,\ref{fig:chirality_promotes_pf}, \ref{fig:chirality_suppresses_oscillations}, and \ref{fig:oscillations_become_stationary} consists of only a single aggregate per species. This is consistent with the results of \citet{potts_painter_1_aggregate} and \citet{yurij_2025_nonlocal}, who showed that, in the absence of growth dynamics, steady-state solutions in a similar nonlocal model with a single species in 1D should have only a single peak. More generally, there is evidence that mass-conserving systems often evolve toward a single aggregate steady-state, even in local models \citep{ishihara2007_single_aggregate_RD, otsuji2007mass_conserving_1_aggregate}. In our models, the final aggregates of the two species can either be separated (\textbf{Fig.\,\ref{fig:chirality_promotes_pf}\,c}, \textbf{e}, \textbf{f}, \textbf{g} and \textbf{Fig.\,\ref{fig:oscillations_become_stationary}}) or mixed (\textbf{Fig.\,\ref{fig:chirality_promotes_pf}} \textbf{d}, \textbf{h}, and \textbf{Fig.\,\ref{fig:chirality_suppresses_oscillations}} \textbf{f}). In the next section, we explore whether this outcome can be predicted by linear theory.

\section{Linear analysis of separation or mixing of species}
\label{sec:linear_analysis_separation_mixing}

Our simulations and linear theory both indicate that chirality can promote stationary pattern formation by reducing the tendency of chase-and-run dynamics to suppress it. This raises an important question: do chase-and-run dynamics have any function beyond simply suppressing pattern formation? In the following sections, we look beyond whether patterns can form, and start to examine their characteristics. For example, in a stationary pattern, are the chasers and runners separated (out of phase) or mixed (in phase)? On zebrafish skin, melanophores are separated from xanthophores, creating distinct black and yellow stripes. Analogously, in predator prey systems, there is a question of when different species are predominantly interspersed or separated, with schooling of fish an example of the latter. How do chase-and-run dynamics influence the outcome of segregation or mixing?

To understand the conditions under which separation or mixing occurs, we can once again apply linear analysis and use the dispersion relation from Equation \eqref{eq:matrix_dispersion}. However, rather than focusing on the eigenvalues (growth rates), $\lambda$, we will look at the corresponding eigenvectors, $\begin{pmatrix} a_\chaser\\ a_\runner\end{pmatrix}$. If $a_\chaser$ and $a_\runner$ have the same sign, then the two species will be in phase within the initial pattern, thereby predicting mixing. Conversely, if $a_\chaser$ and $a_\runner$ have opposite sign, then the two species will be out of phase, predicting separation \citep{murray_book_chapter_6}.

The conditions can be found by rearranging Equations \eqref{eq:matrix_dispersion} and \eqref{eq:lambda} to give
\begin{align}
        a_{\runner} &= \frac{(1-D)k^2+(\Lambda_{\runner\runner}-\Lambda_{\chaser\chaser}) +A}{2\Lambda_{\chaser\runner}} \, a_\chaser, \label{eq:chiral_a_v_a_u}\\
        a_{\chaser} &= \frac{(D-1)k^2+(\Lambda_{\chaser\chaser}-\Lambda_{\runner\runner}) + A}{2\Lambda_{\runner\chaser}} \, a_\runner, \label{eq:chiral_a_u_a_v}
\end{align}
where $A \equiv \sqrt{4\Lambda_{\chaser \runner}\Lambda_{\runner \chaser}+[(1-D)k^2+(\Lambda_{\runner \runner}-\Lambda_{\chaser \chaser})]^2}\geq 0$ is taken as the positive root, as this will correspond to the larger $\lambda$ and faster growing perturbation. Since we focus on stationary patterns, we assume that $\lambda\in\mathbb{R}$, which requires sufficiently large differential self-interaction, as discussed in Section \ref{sec:insights_PF}. For complex $\lambda$, the resulting pattern would exhibit spatio-temporal oscillations, with the phase difference between chasers and runners oscillating over time, and so whether they are separated or mixed would change over time.

In the stationary case, using the fact that $\Lambda_{\chaser\runner}>0$ and $\Lambda_{\runner\chaser}<0$, and that $A\geq 0$, we can infer the following conditions on the phase relationship:

\vspace{0.3cm}
\begin{boxx}
If $\Lambda_{\runner\runner}-Dk^2>\Lambda_{\chaser\chaser}-k^2$, then by Equation \eqref{eq:chiral_a_v_a_u}, $\chaser$ and $\runner$ are in phase.
\\
If $\Lambda_{\runner\runner}-Dk^2<\Lambda_{\chaser\chaser}-k^2$, then by Equation \eqref{eq:chiral_a_u_a_v}, $\chaser$ and $\runner$ are out of phase.
\\ \\
In other words, \textbf{according to linear theory}, if the runners self-aggregate faster than the chasers, then the two species will be separated. Conversely, if the chasers self-aggregate faster than the runners, then the two species will be mixed.
\\ \\
Note that $\Lambda_{\runner\runner}-Dk^2$ and $\Lambda_{\chaser\chaser}-k^2$ are measures of the self-aggregation of runners and chasers, respectively, at wavenumber $k$. For each species, the self-aggregation is increased by attractive nonlocal self-interactions and decreased by repulsive nonlocal self-interactions and diffusion.
\label{boxx:phase_conditions}
\end{boxx}
\vspace{0.3cm}

We can build a rough intuitive understanding of this prediction as follows: the species that aggregates more slowly will form its pattern in response to the already-established pattern of the faster-aggregating species. This means that the slower aggregator dictates whether the two species are mixed or separated. If chasers aggregate more slowly, they will move towards the pre-formed clusters of runners, leading to a mixed state. Conversely, if runners aggregate more slowly, they will move away from the already-formed chaser clusters, resulting in separation.

This prediction implies that the magnitude of cross-interactions have no effect on the outcome of mixing or separation, which is arguably counter-intuitive. For example, it implies that even an extremely high chasing strength, $\mu_{\chaser\runner}$, would not lead to the chasers `catching' the runners and the two species mixing. Being cautious of the limitations of linear theory \citep{krause2024turing_conditions_not_enough, oliver2024_turing_arent_enough}, we will check these predictions against simulations.

\begin{figure}
    \centering
    \includegraphics[width=0.97\textwidth]{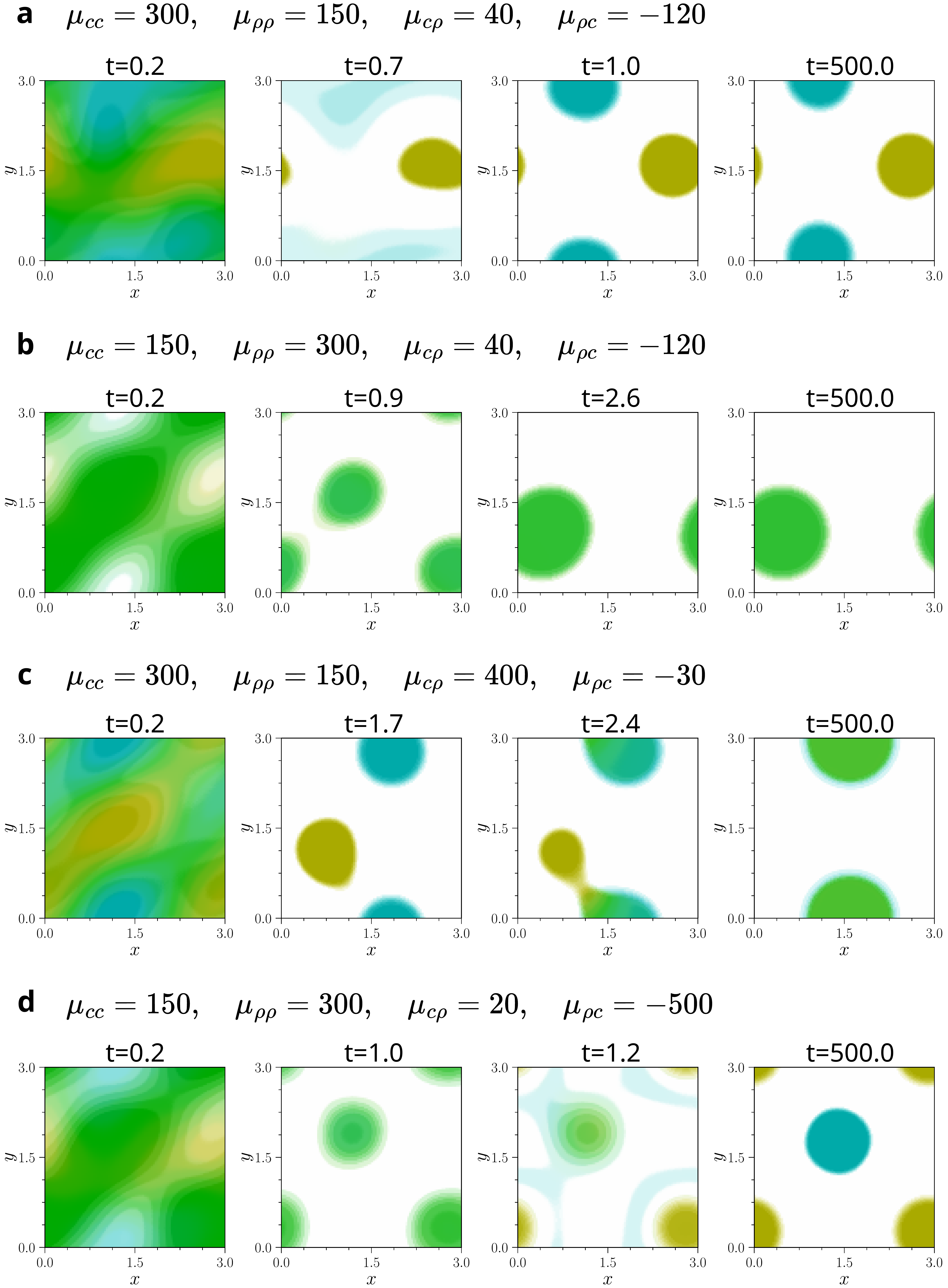}
    \caption{Simulations of chase-and-run with self-attracting chasers and runners. In \textbf{a} and \textbf{c}, stronger chaser self-attraction ($\mu_{\chaser\chaser}>\mu_{\runner\runner}$) leads to species separation at early times, as predicted by linear theory. This persists in \textbf{a} but not in \textbf{c}, where cross-species chasing is much higher than running ($|\mu_{\chaser\runner}|\gg|\mu_{\runner\chaser}|$), leading to mixing \changes{that starts to appear around $t=2.4$ and is maintained into late times.} Conversely, in \textbf{b} and \textbf{d}, stronger runner self-attraction ($\mu_{\runner\runner}>\mu_{\chaser\chaser}$) causes mixing at early times. This persists in \textbf{b} but not in \textbf{d}, where cross-species running is much higher than chasing ($|\mu_{\runner\chaser}|\gg|\mu_{\chaser\runner}|$), leading to separation at \changes{around $t=1.2$} and a separated state at late times. The colourmap from \textbf{Fig.\,\ref{fig:chirality_promotes_pf}}\,\textbf{i} is used throughout, with colours rescaled only for $t=0.2$ heatmaps to display amplitudes of order $\mathcal{O}(10^{-3})$. All simulations use the direct sensing model, although gradient sensing simulations show the same qualitative behaviour. In all cases, $\alpha_{\chaser \runner}=-20^\circ$, $\alpha_{\chaser \runner}=70^\circ$; other parameters are given in Table \ref{tab:parameters} of Appendix \ref{sec:appendix_parameter_values}. Videos of these simulations are available at \citet{My_GitHub}.}  \label{fig:separation_vs_mixing}
\end{figure}

\section{Simulations of separation or mixing of species}
\label{sec:simulations_separation_mixing}

\subsection{Chase-and-run}
The linear analysis in Section \ref{sec:linear_analysis_separation_mixing} provides predictions for whether chasers and runners will mix or separate in stationary patterns, based on their relative self-aggregation strengths. To test these predictions, we perform simulations under various conditions and examine the resulting phase relationships between species.

In all cases, we find that the phase relationships predicted by linear theory hold at early times, before nonlinear effects become significant. Furthermore, when one species is self-repulsive, the predictions also seem to remain valid even into later times. For example, in \textbf{Fig.\,\ref{fig:chirality_promotes_pf}}\,\textbf{d} and \textbf{h}, the chasers self-repel, and thus aggregate slower than the runners, which leads to mixing. In contrast, for \textbf{Fig.\,\ref{fig:chirality_promotes_pf}}\,\textbf{e}, the runners self-repel, leading to separation. 

When both species self-attract, the story is more complicated. In \textbf{Fig.\,\ref{fig:chirality_promotes_pf}}\,\textbf{c}, \textbf{f}, and \textbf{g}, both species are self-attractive, but runners have a higher diffusivity (and therefore aggregate more slowly), resulting in species separation at early times that persists until late times. This behaviour aligns with the predictions from Section \ref{sec:linear_analysis_separation_mixing}. However, when the \textit{magnitude} of cross-interactions are sufficiently asymmetric --- i.e. the chasing, $|\mu_{\chaser \runner}|$, is much faster than the running, $|\mu_{\runner\chaser}|$, or vice versa --- we see that phase relationships can change at later times, as shown in \textbf{Fig.\,\ref{fig:separation_vs_mixing}}.

In \textbf{Fig.\,\ref{fig:separation_vs_mixing}}, at early times, all simulations follow the expected trends: in \textbf{a} and \textbf{c}, stronger chaser self-attraction ($\mu_{\chaser\chaser}>\mu_{\runner\runner}$) leads to separation, whilst in \textbf{b} and \textbf{d}, stronger runner self-attraction ($\mu_{\runner\runner}>\mu_{\chaser\chaser}$) leads to mixing. However, at later times, in \textbf{c} and \textbf{d}, the phase relationship changes. In \textbf{Fig.\,\ref{fig:separation_vs_mixing}}\,\textbf{c}, although chasers still self-attract more strongly, the cross-species chasing strength, $|\mu_{\chaser \runner}|$, is much larger than the running strength, $|\mu_{\runner \chaser}|$. This enables the chasers to effectively ‘catch’ the runners, when pattern amplitudes are sufficiently large, ultimately resulting in a mixed state. Conversely, in \textbf{Fig.\,\ref{fig:separation_vs_mixing}}\,\textbf{d}, where running is much stronger than chasing, the runners eventually ‘escape’ from the chasers, leading to a final separated state despite initially being mixed.

We conjecture that mixed states are more robust to nonlinear effects than separated states. Anecdotally, we observe in our simulations that systems which linear theory predicts to be mixed typically require a much greater excess of running over chasing strength to become separated than the excess of chasing over running strength needed to drive mixing from a separated state. We speculate on two possible explanations. First, volume-filling may promote mixing, as discussed in the next section. Second, in a mixed configuration, where a runner spot sits atop a chaser spot, the radial outward ‘force’ from running must overcome the runner’s self-attraction, which acts radially inward. In contrast, in a separated state, chasers feel a directed attraction toward a nearby runner spot without any comparable opposing force, making mixing potentially easier to initiate. A systematic investigation of this phenomenon is beyond the scope of this study and left for future work. Understanding when and why structures laid down in the linear regime persist into the nonlinear regime would help clarify when linear theory remains predictive.

In summary, we find that whilst linear theory often correctly predicts whether species will mix or separate, it underestimates the influence of cross-species interactions. As a result, when these interactions are sufficiently strong or imbalanced, they can drive outcomes on longer timescales that the linear theory fails to predict. In the next section, we go a step further, examining how a mechanism that plays almost no role in the linear theory can nonetheless have a powerful impact on species mixing and separation.

\begin{figure}
    \centering
    \includegraphics[width=1\textwidth]{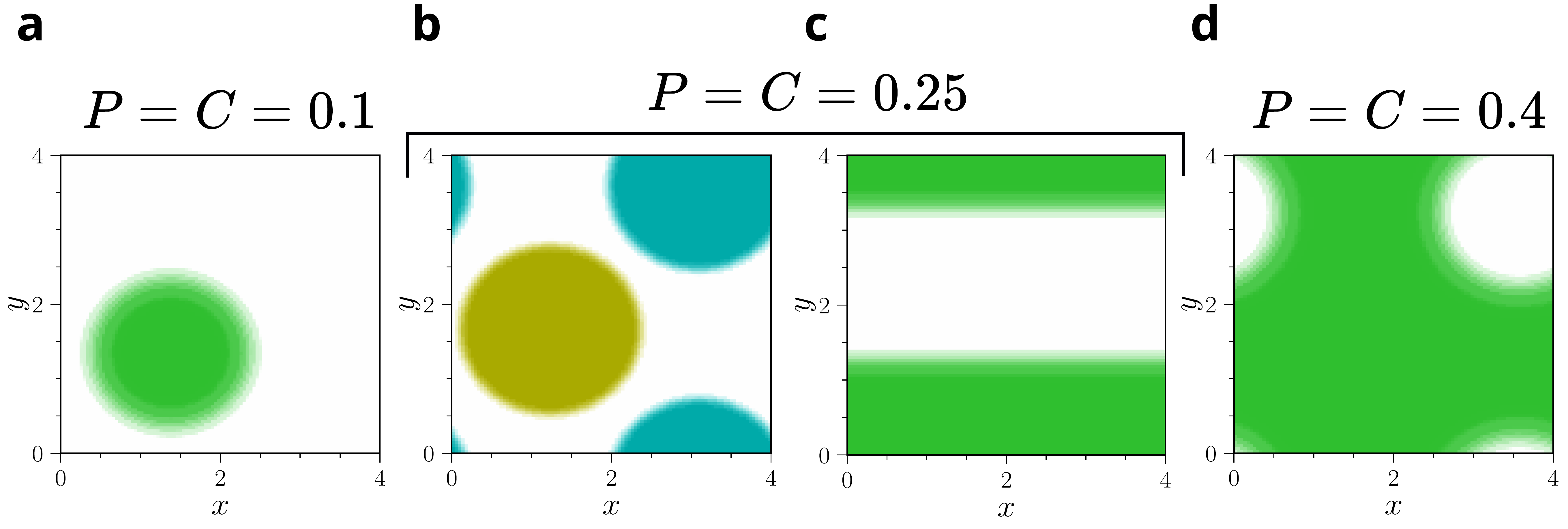}
    \caption{Late-time stationary patterns of simulations of two species with zero cross-interactions and parallel attractive self-interactions, for a range of initial densities, $\Runner$ and $\Chaser$. The two species share all the same parameters and are essentially identical. As we might expect, at early times, the distributions of the two species did not look obviously correlated. However, despite interacting only through volume-filling, the two species can ultimately mix, leading to an in phase pattern at late times for \textbf{a}, \textbf{c}, and \textbf{d}. For each set of parameters, we ran simulations with 5 different realisations of the Gaussian noise initial conditions. We observed the same qualitative behaviour for every realisation when $\Runner=\Chaser=0.1$ and $\Runner=\Chaser=0.4$ (used in \textbf{a} and \textbf{d}, respectively). However, for $\Runner=\Chaser=0.25$ (\textbf{b} and \textbf{c}), 2 realisations lead to separated spots (\textbf{b}) and 3 to a mixed stripe (\textbf{c}). The colourmap from \textbf{Fig.\,\ref{fig:chirality_promotes_pf}}\,\textbf{i} is used throughout. All simulations use the direct sensing model, although gradient sensing simulations exhibit the same behaviour. These systems are nonchiral, with $\alpha_{\chaser\chaser}=\alpha_{\runner\runner}=\alpha_{\chaser\runner}=\alpha_{\runner\chaser}=0$; other parameters are given in Table\,\ref{tab:parameters} of Appendix \ref{sec:appendix_parameter_values}. Videos of these simulations are available at \citet{My_GitHub}.}  \label{fig:volume_filling}
\end{figure}

\subsection{Volume-filling can drive mixing}
Another arguably counter-intuitive phenomenon regarding separation and mixing is observed when considering two species that have zero cross-interactions ($\mu_{\runner \chaser}=\mu_{\chaser\runner}=0$). In such cases, the only influence that one species can have on the other is through volume-filling effects, which are present through the terms $\changes{\phi}_\chaser(\chaser, \runner)=\changes{\phi}_\runner(\chaser, \runner)=1-\chaser-\runner$ in Equations \eqref{eq:two_species_direct} and \eqref{eq:two_species_gradient}. The linear theory in Section \ref{sec:linear_analysis_separation_mixing} predicts that having zero cross-interaction would lead to the location of one species being uncorrelated with the location of the other. We might naively guess that  volume-filling, which is a purely nonlinear effect, could introduce bias towards species separation.  Instead, we see simulations in which the two species become perfectly in phase and mixed in the late-time pattern, as shown in \textbf{Fig.\,\ref{fig:volume_filling}}. 

For systems with small initial densities (\textbf{Fig.\,\ref{fig:volume_filling}}\,\textbf{a}) and high initial densities (\textbf{Fig.\,\ref{fig:volume_filling}}\,\textbf{d}), we always observed mixing. Curiously, for intermediate initial densities, some realisations of the Gaussian noise initial conditions lead to mixing (\textbf{Fig.\,\ref{fig:volume_filling}}\,\textbf{c}), whilst other realisations lead to separation (\textbf{Fig.\,\ref{fig:volume_filling}}\,\textbf{b}).  As is the case with phenomena described in other sections, linear theory is consistently correct at early times: the species distributions from the simulations in \textbf{Fig.\,\ref{fig:volume_filling}} appear uncorrelated at early times. They only become in phase later, when pattern amplitudes are larger.

A related phenomenon was observed by \citet{bailo_carrillo_2023_freezing} in a model of nonlocal two-species cell-cell adhesion. At densities near carrying capacity, volume-filling greatly reduces mobility, and they showed this could effectively `freeze' a species-mixed initial condition, preventing other dynamics from driving separation. In our simulations in \textbf{Fig.\,\ref{fig:volume_filling}}, however, volume-filling seems to play an even more active role, not just maintaining mixing, but actively driving the two species into phase alignment. One conjectured explanation is that where both species happen to overlap, the total density will be closer to carrying capacity, reducing local mobility. Meanwhile, individuals from lower-density regions, which are still mobile, could be drawn towards these locations by the attractive interactions. Since motion slows only upon arrival, this could create a self-reinforcing accumulation of both species in the same location, gradually promoting in phase alignment and a mixed state.

A final remark on \textbf{Fig.\,\ref{fig:volume_filling}} is that it also demonstrates a pattern transition as the initial density increases: from spots to stripes to holes. The initial densities at which this transition occurs is dependent on domain size.

A more comprehensive investigation into the precise ways that volume-filling influences species separation and mixing is left for future work. What emerges clearly from our results, however, is that volume-filling highlights a notable limitation of linear theory in predicting the spatial co-location of two species. Unlike cross-species interactions, whose effects are present but underestimated, volume-filling appears in the linear analysis only as a constant prefactor and plays virtually no dynamic role. Yet despite this minimal role in linear theory, it can drive species-mixing and significantly reshape the long-term behaviour of these systems.

\section{Travelling pulses}
\label{sec:other_complex_behaviour}

\subsection{Population-level chase-and-run}
In this final section, we explore how chirality can influence travelling pulse behaviour in the models. One such form of travelling pulse is \textit{population-level} chase-and-run, a spatio-temporal pattern where aggregates of chasers continually pursue aggregates of runners across the domain. This behaviour was recently studied by \citet{painter_giunta_2024_chase_run}, who showed that it occurs robustly when chasers have a sufficiently larger interaction range for sensing runners, $\xi_{\chaser \runner}$, than runners do for sensing chasers, $\xi_{\runner \chaser}$. As with the other patterns investigated in our work, these pulses can emerge spontaneously from small heterogeneous perturbations about a homogeneous state. In their study, \citet{painter_giunta_2024_chase_run} used a nonlocal direct sensing model, similar to Equation\,\eqref{eq:two_species_direct}, with a tophat kernel in one and two dimensions.

\begin{figure}
    \centering
    \includegraphics[width=1\textwidth]{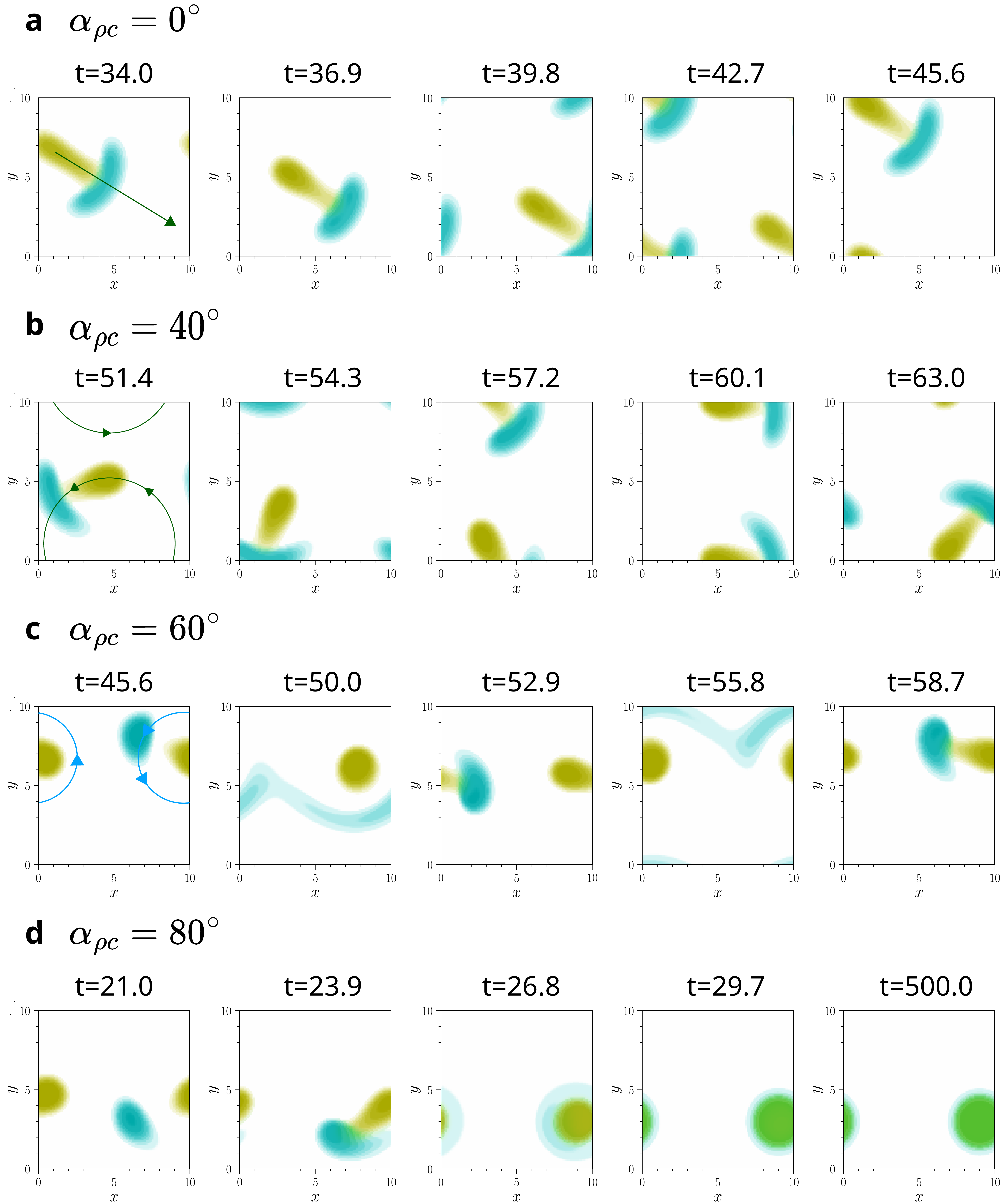}
    \caption{Simulations with population chase-and-run with chirality. When cross-species running has a shorter interaction range than cross-species chasing ($\xi_{\runner\chaser}<\xi_{\chaser\runner}$), chasers form an aggregate that continually pursues an aggregate of runners \citep{painter_giunta_2024_chase_run}. Here, $\xi_{\chaser \chaser}=\xi_{\runner \runner}=\xi_{\chaser \runner}=2$, whilst $\,\xi_{\runner \chaser}=1$. \textbf{a} Without chirality ($\alpha_{\runner \chaser}=0$) this pulse travels in a straight line. \textbf{b} Introducing chirality ($\alpha_{\runner \chaser}=40^\circ$) induces population chase-and-run along a curved path. \textbf{c} Larger angles can cause the runner aggregate to orbit the chasers, which move less. \textbf{d} Even larger angles lead to chasers `catching' runners, yielding a stationary mixed state. Arrows in the first panel of \textbf{a}-\textbf{c} roughly indicate the direction of movement. The colourmap from \textbf{Fig.\,\ref{fig:chirality_promotes_pf}}\,\textbf{i} is used. All simulations shown use the direct sensing model, although simulations with gradient sensing yield the same qualitative behaviour. In all cases, $\alpha_{\chaser \chaser}=\alpha_{\runner \runner}=\alpha_{\chaser \runner}=0^\circ$; other parameters are given in Table \ref{tab:parameters} of Appendix \ref{sec:appendix_parameter_values}. Videos of these simulations are available at \citet{My_GitHub}.}
    \label{fig:population_chase_run}
\end{figure}

Our simulations demonstrate that their conclusions are robust to some changes in model details. We find similar population-level chase-and-run behaviour in the gradient sensing model and also when using exponential kernels. Some examples are shown in \textbf{Fig.\,\ref{fig:population_chase_run}}. However, we also observe that this behaviour breaks down if the initial densities, $\Runner$ and $\Chaser$, are sufficiently high that the resulting aggregates are sufficiently large to self-interact across the periodic boundary. In these cases, the ordered population-level chase-and-run breaks down into more complex spatio-temporal behaviour. That said, it is not clear that this limitation would still apply on an infinite domain or with non-periodic boundaries.

Furthermore, we find that chirality, introduced through a non-zero running angle, can significantly alter the nature of population-level chase-and-run. For moderate angles, as in \textbf{Fig.\,\ref{fig:population_chase_run}}\,\textbf{b}, the chaser and runner aggregates follow curved, rather than straight, trajectories. For larger angles, such as in \textbf{c}, the runner aggregate turns more sharply and can effectively orbit the chaser aggregate, which shifts its position less. When the angle becomes sufficiently large, as in \textbf{d}, the runners can no longer escape quickly enough. The chasers then catch the runners, resulting in a stationary, species-mixed pattern. This is consistent with our earlier findings in Sections \ref{sec:insights_PF} and \ref{sec:simulations_1}, where we showed that chirality can suppress spatio-temporal behaviour and promote the formation of stationary structures.

We have only observed population-level chase-and-run behaviour emerging from systems with a Turing-wave bifurcation, that is, when $\text{Re}[\lambda(k)]>0$ and $\text{Im}[\lambda(k)]\neq 0$, $k>0$. However, the converse does not hold: in \textbf{Fig.\,\ref{fig:population_chase_run}}\,\textbf{d}, the pattern becomes stationary due to the large angle $\alpha_{\runner \chaser}=80^\circ$, and although the imaginary component of the growth rate is reduced compared to the non-chiral case, it is still non-zero. A sufficiently larger angle would indeed eliminate this imaginary component, resulting in a standard Turing bifurcation, and likely a stationary pattern. However, \ref{fig:population_chase_run}\,\textbf{d} shows that the transition from population-level chase-and-run to stationary pattern can occur at smaller angles than this bifurcation point. This highlights a nonlinear influence of chirality on patterning.

\begin{figure}
    \centering
    \includegraphics[width=1\textwidth]{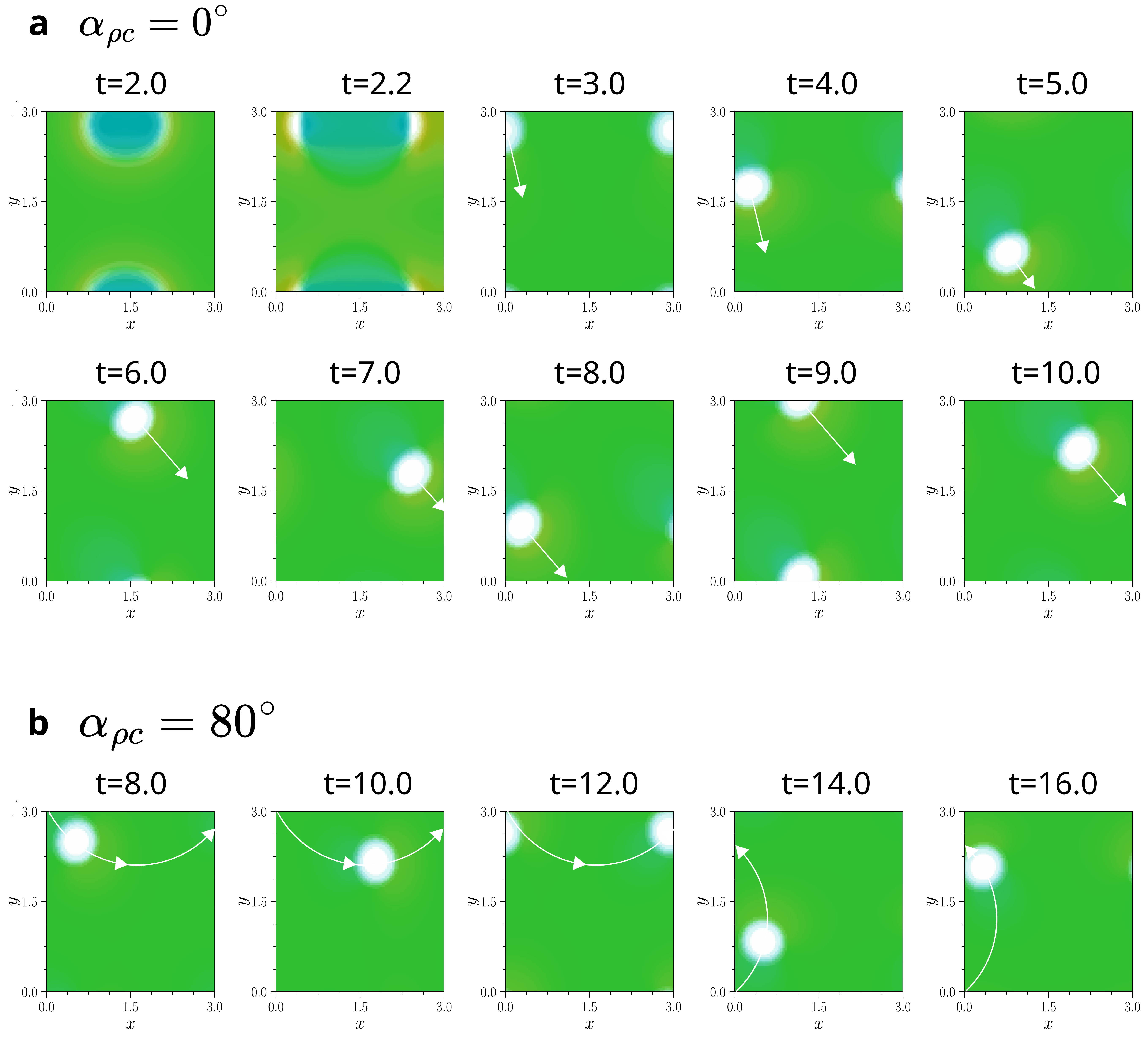}
    \caption{Two simulations (\textbf{a},\textbf{b}) exhibiting travelling holes. In both cases, a single runner spot initially forms, encircled by both species and separated by a thin low-density gap in both populations (see first two panels of \textbf{a}). This gap then merges into a single hole that moves across the domain. After some transient movement: in the non-chiral case (\textbf{a}), the hole moves in a straight line; with chirality (\textbf{b}), it follows a curved trajectory. In both cases, this motion persist into late times. There is a higher density of chasers at the front of the hole (in the direction of motion), and a higher density of runners at the back. The colourmap from \textbf{Fig.\,\ref{fig:chirality_promotes_pf}}\,\textbf{i} is used throughout. In both cases, the initial density is near carry capacity ($P=C=0.48$), and $\alpha_{\chaser \chaser}=\alpha_{\runner \runner}=\alpha_{\chaser \runner}=0^\circ$; other parameters are given in Table \ref{tab:parameters} of Appendix \ref{sec:appendix_parameter_values}. Videos of these simulations are available at \citet{My_GitHub}.}  \label{fig:travelling_holes}
\end{figure}

\subsection{Travelling holes}

As well as population-level chase-and-run, these models can also produce another form of travelling pulse, which we refer to as `travelling holes'. These holes are localised regions of near-zero density for both species surrounded by a background of high, nearly uniform density.  Like population-level chase-and-run, these patterns are driven by chase-and-run interactions and the holes can travel persistently across the domain.

In contrast to population-level chase-and-run, we observe travelling holes arising in systems where the linear growth rate $\lambda(k)$ is purely real, meaning that linear analysis predicts a standard Turing bifurcation and the emergence of stationary patterns. Their existence shows that nonlinear effects can generate spatio-temporal oscillations that are not predicted by linear analysis. Whilst previous studies have shown examples of models where linear theory fails to predict oscillations \citep{painter2011_keller_segel_chaos, Painter2015nonlocal}, those oscillations were driven by reaction kinetics. In contrast, the travelling holes observed in our work emerge solely from movement dynamics.

Unlike population-level chase-and-run pulses, travelling holes do not require asymmetry between interaction ranges. We observe them even when all interaction ranges are equal. They are observed to emerge when the magnitude of the chasing strength, $|\mu_{\chaser \runner}|$, is much larger than that of the running strength, $|\mu_{\runner \chaser}|$, and when the initial densities are close to carrying capacity. However, the precise conditions under which these patterns form remain unclear and are not within the scope of the present study.

\textbf{Fig.\,\ref{fig:travelling_holes}} shows examples of travelling holes. The structure of the holes is heterogeneous: there is a higher density of chasers at the front, in the direction of motion, and a higher density of runners at the back. This structure reflects the underlying dynamics, with chasers pursuing populations ahead and runners moving away from populations behind. 

In both the simulations in \textbf{Fig.\,\ref{fig:travelling_holes}}, an initial pattern forms in which a spot of runners appears, surrounded by a background of chasers and runners. This is consistent with the linear prediction that species will separate when the chaser self-attraction is stronger than the runner self-attraction (see Section~\ref{sec:linear_analysis_separation_mixing}). A thin low-density gap separates the runner spot from the surrounding chasers. As nonlinear effects take over, this gap merges into a single, low-density hole which then travels across the domain, settling into a straight-line trajectory, when chirality is not present. With a non-zero running angle, and thus chirality, the holes still form, but their paths become curved rather than straight, as shown in Fig\,\ref{fig:travelling_holes}\,\textbf{b}. Therefore, for both travelling holes and population-level chase-and-run, chirality at the individual level leads to chiral movement at the population and aggregate level.

\section{Discussion}
\label{sec:discussion}

Chase-and-run dynamics are observed across many biological systems, from predator-prey interactions in ecology \citep{white1998_elk_wolves, ripple2000_elk_wolves,  gaynor2019landscape_of_fear} to cellular systems in biology \citep{theveneau2013_chase_run_neural_crest, szabo2015_chase_run_common}. In some cases, these dynamics involve chiral movement such as in zebrafish pigment cells \citep{YamanakaKondo2014vitro}, or in prey species that exhibit behavioural lateralisation, evading predators by moving away at an angle \citep{miletto2020_lateralisation_well_cited_review, gobbo2025_zebrafish_behaviour_lateralisation_mini_review, cantalupo1995_population_level_lateralization}. In this work, we have explored how chirality can affect pattern formation in chase-and-run systems by generalising nonlocal advection-diffusion models, including both direct and gradient sensing frameworks, to allow for movement at any angle relative to the separation of the interacting populations.

Through linear stability analysis and numerical simulations, we found that chirality can have a substantial impact on the capacity for pattern formation. Because the perpendicular component of motion does not affect the linear stability of a heterogeneous perturbation, the impact of any interaction on the capacity for pattern formation can be reduced by increasing the angle of motion. In this way, introducing chirality is predicted, and has been numerically observed, to inhibit the pattern-suppressing effects of chase-and-run interactions. This can enable systems with chiral chase-and-run interactions to form patterns where equivalent systems with parallel chase-and-run cannot. Similarly, chirality can suppress one mechanism of generating spatio-temporal oscillations, and thereby promote stationary patterns. The shapes of these stationary patterns do not necessarily appear chiral themselves, in that they are left-right symmetric. Overall, our results support the hypothesis of \citet{YamanakaKondo2021_chirality_function} that chirality has the potential to play roles in cellular systems beyond breaking left-right symmetry in emergent structures. 

That said, we have observed examples of spatio-temporal behaviour in numerical simulations where chiral interactions at the individual level resulted in chiral movement at the population level. For example, the population-level chase-and-run pulses studied by \citet{painter_giunta_2024_chase_run} and the travelling hole patterns observed in our simulations both followed curved paths when chirality was introduced. These findings suggest that organisms that exhibit intrinsic chirality may form structures that also move chirally at the population-level.

Our linear stability analysis reliably predicted early-time behaviour, and although it remained accurate into late-times in many cases, there are some notable exceptions. It accurately identified the conditions for pattern formation from small perturbations about a homogeneous state. However, it sometimes failed to predict spatio-temporal oscillations at later times: we saw examples showing that the linear growth rate having a non-zero imaginary component is neither a sufficient nor a necessary condition for spatio-temporal oscillations at late times.

Linear analysis provided simple conditions for whether runners and chasers would be mixed or separated in a stationary pattern: if the runners self-aggregate faster than the chasers, then the two are predicted to be separated, and if the converse is true, then they will be mixed. These predictions were accurate into late-times for our simulations, other than in cases where both species were self-attracting and there was a large difference in the magnitude of the cross-interactions. Surprisingly, we also saw that volume-filling, despite having little effect in linear theory, can drive in phase mixing. Although there is no clear \textit{a priori} reason why linear analysis should successfully be able to predict separation or mixing in the nonlinear regime, the fact that it fails may still be unexpected given that it typically succeeds in more commonly studied local Turing models \citep{woolley_krause_gaffney2021bespoke_turing}. This highlights the importance of studying more complex models, like these nonlocal models, to test whether or not assumptions inherited from simpler frameworks apply in more complex biological systems.

The above limitations of the linear analysis may potentially point towards subcritical bifurcations in these systems. In systems with subcriticality, dynamics can jump to a high-amplitude solution branch, and these solutions may exhibit behaviour that deviates significantly from linear predictions\,\,\citep{krause2024_subcriticality}. We have implemented preliminary simulations varying a model parameter near a Turing bifurcation point, and seen little change in the amplitude of the pattern produced, which indeed suggests a subcritical bifurcation. Additionally, \citet{giunta2024weakly_nonlinear} and \citet{yurij_2025_nonlocal} have shown analytically that similar nonlocal models can exhibit subcritical bifurcations.

Our results emphasise the need for analytical tools which go beyond local linear stability analyses in order to understand complex biological systems in the long-term. Approaches such as weakly-nonlinear analysis \citep{giunta2024weakly_nonlinear} and Crandall-Rabinowitz analysis \citep{yurij_2025_nonlocal} can provide further insights, such as deducing the bifurcation structure in these systems, and characterising solution branches, but only in regions of the parameter space close to bifurcations points. Energy minimisation methods \citep{giunta2022_energy, potts2024_heterogeneous} are also powerful tools for identifying steady-states, even far from bifurcations, though their application depends on the ability to define an appropriate energy functional --- something that is not always possible for these models.

A natural question arising from our results is how the perpendicular component of motion affects dynamics beyond the linear regime. Whilst it has no impact on linear instability, our simulations show that it clearly influences the nonlinear evolution of the system. The influence of chirality at the weakly nonlinear level could be explored, as well as whether chirality changes the bifurcation structure of these systems.

Another valuable extension to this work would be to include more biological details, in order to explore a specific biological system. For example, in zebrafish, although it has been shown that the interactions between melanophores and xanthophores alone are sufficient for the formation of stripes on fins and spots on bodies \citep{kondo2021_zebrafish_review, mahalwar2014_iridophores_maybe_needed, WatanabeKondo2015_iridophores_not_needed, volkening2020_fins_ABM}, \citet{volkening2018iridophores_ABM} have shown that a third pigment cell type, iridophores, contributes to \textit{robust} stripe formation on the body. Moreover, there exist multiple sub-types of each pigment cell type, whose different interactions may all be important for the specific details of the pattern. This suggests that insight may be gained by extending our models to three or more species and also considering reaction-kinetics, as has already been implemented in ABM models \citep{owen2020YatesZebrafishABM, volkening2018iridophores_ABM}.

As mathematical models grow more faithful to specific biological systems, we face a trade-off between incorporating detail and maintaining interpretability. It becomes increasingly important to identify which features are vital to retain. Exploring simpler, more general models, such as in this study, can help with this. Our work suggests that chirality is indeed an important feature, influencing pattern formation in unexpected ways.

\backmatter

\bmhead{Acknowledgements} We gratefully acknowledge Professor José Carrillo for his insight on the link between direct sensing and gradient sensing models.
\,Thomas Jun Jewell is supported by funding from the Edmund J. Crampin Scholarship at Linacre College, University of Oxford; and the Engineering and Physical Sciences Research Council [EP/W524311/1].

\section*{Declarations}
The authors have no competing interests to declare.

\bmhead{Rights Retention Statement}
For the purpose of Open Access,  a CCBY public copyright licence is applied to any Author Accepted Manuscript version arising from this submission.

\section*{Data availability}
All code and assosciated documentation can be found at the Github repository, \url{https://github.com/JunJewell/ChiralNonlocal}

\begin{appendices}
\section{Parameter values}
\label{sec:appendix_parameter_values}
\noindent
\begin{tabular}{|c|c|c|c|c|c|c|c|c|c|c|c|}
    \hline
    \textbf{Figure} & \textbf{Model Type} & \textbf{Kernel} & $\Chaser$ & $\Runner$ & $D$ & all $\xi$ & $\mu_{\chaser\chaser}$ & $\mu_{\chaser\runner}$ & $\mu_{\runner\chaser}$ & $\mu_{\runner\runner}$ & \changes{L} \\
    \hline
    \ref{fig:phase_map} & \changes{direct} & \changes{tophat} & \changes{0.25} & \changes{0.25} & \changes{(0,5)} & \changes{1} & \changes{50} & \changes{50} & \changes{-50} & \changes{50} & \changes{8}\\
    \ref{fig:chirality_promotes_pf}\,\textbf{c} & direct & tophat & 0.25 & 0.25 & 4 & 1 & 50 & 50 & -50 & 50 & \changes{8}\\
    \ref{fig:chirality_promotes_pf}\,\textbf{d} & direct & tophat & 0.25 & 0.25 & 1 & 1 & -50 & 50 & -50 & 50 & \changes{8}\\
    \ref{fig:chirality_promotes_pf}\,\textbf{e} & direct & tophat & 0.25 & 0.25 & 1 & 1 & 50 & 50 & -50 & -50 & \changes{8}\\
    \ref{fig:chirality_promotes_pf}\,\textbf{f} & gradient & exponential & 0.25 & 0.25 & 4 & 0.5 & 40 & 40 & -40 & 40 & \changes{8}\\
    \ref{fig:chirality_promotes_pf}\,\textbf{g} & gradient & tophat & 0.25 & 0.25 & 4 & 1 & 20 & 20 & -20 & 20 & \changes{8}\\
    \ref{fig:chirality_promotes_pf}\,\textbf{h} & gradient & tophat & 0.25 & 0.25 & 1 & 1 & -20 & 20 & -20 & 20 & \changes{8}\\
    \ref{fig:chirality_suppresses_oscillations} i \& \ref{fig:oscillations_become_stationary} & direct & tophat & 0.25 & 0.25 & 1 & 1 & 20 & 20 & -20 & 50 & \changes{8} \\
    \ref{fig:chirality_suppresses_oscillations} ii & direct & tophat & 0.25 & 0.25 & 1 & 1 & 50 & 20 & -20 & 20 & \changes{8}\\
    \ref{fig:chirality_suppresses_oscillations} iii & gradient & tophat & 0.25 & 0.25 & 1 & 1 & 8 & 8 & -8 & 20 & \changes{8}\\
    \ref{fig:chirality_suppresses_oscillations} iv & gradient & tophat & 0.25 & 0.25 & 1 & 1 & 20 & 8 & -8 & 8 & \changes{8}\\
    \ref{fig:chirality_suppresses_oscillations} v & direct & exponential & 0.25 & 0.25 & 1 & 0.5 & 20 & 20 & -20 & 50 & \changes{8}\\
    \ref{fig:chirality_suppresses_oscillations} vi & direct & exponential & 0.25 & 0.25 & 1 & 0.5 & 50 & 20 & -20 & 20 & \changes{8}\\
    \ref{fig:separation_vs_mixing} \textbf{a} & direct & exponential & 0.1 & 0.1 & 1 & 0.4 & 300 & 40 & -120 & 150 & \changes{3}\\
    \ref{fig:separation_vs_mixing} \textbf{b} & direct & exponential & 0.1 & 0.1 & 1 & 0.4 & 150 & 40 & -120 & 300 & \changes{3}\\
    \ref{fig:separation_vs_mixing} \textbf{c} & direct & exponential & 0.1 & 0.1 & 1 & 0.4 & 300 & 400 & -30 & 150 & \changes{3}\\
    \ref{fig:separation_vs_mixing} \textbf{d} & direct & exponential & 0.1 & 0.1 & 1 & 0.4 & 150 & 20 & -500 & 300 & \changes{3}\\
    \ref{fig:volume_filling} \textbf{a} & direct & exponential & 0.1 & 0.1 & 1 & 0.4 & 100 & 0 & 0 & 100 & \changes{4}\\
    \ref{fig:volume_filling} \textbf{b} \& \textbf{c} & direct & exponential & 0.25 & 0.25 & 1 & 0.4 & 100 & 0 & 0 & 100 & \changes{4}\\
    \ref{fig:volume_filling} \textbf{d} & direct & exponential & 0.4 & 0.4 & 1 & 0.4 & 100 & 0 & 0 & 100 & \changes{4}\\
    \ref{fig:population_chase_run} \textbf{a-d} & direct & exponential & 0.05 & 0.05 & 1 & see caption & 100 & 100 & -100 & 100 & \changes{10}\\
    \ref{fig:travelling_holes} & direct & exponential & 0.48 & 0.48 & 1 & 0.4 & 600 & 400 & -30 & 300 & \changes{3}\\
    \hline
\end{tabular}
\captionof{table}{Parameters used for the dispersion curves and simulations in each figure.}
\label{tab:parameters}

\section{\changes{Hankel transform derivation}}
\label{sec:appendix_hankel}

\changes{Here we show the derivation of the Hankel transform in the dispersion relation, illustrating the steps between the first and second line of Equation\,\eqref{eq:dispersion_single_species_direct}. The first line of this equation is
\begin{equation}
\begin{split}
    \lambda(\boldsymbol{k}) &= -k^2 -\Runner \changes{\phi}(\Runner)\frac{\partial g}{\partial \runner}\Big\rvert_{\Runner}\,\frac{\mu}{\xi^2}\,ik\boldsymbol{\hat{k}}\cdot \boldsymbol{\underline{\underline{R}}}(\alpha)\int_{\mathbb{R}^2} \boldsymbol{\hat{s}}\,\Omega\left(\frac{s}{\xi}\right)\, e^{i\boldsymbol{k}\cdot\boldsymbol{s}} \diff^2 \boldsymbol{s},
\end{split}
\label{eq:appendix_dispersion_first_line}
\end{equation}
where, as a reminder, $s \equiv |\boldsymbol{s}|$ and $k \equiv |\boldsymbol{k}|$, with $\boldsymbol{\hat{s}}\equiv\frac{\boldsymbol{s}}{s}$ and $\boldsymbol{\hat{k}}\equiv\frac{\boldsymbol{k}}{k}$.}

\changes{To evaluate the integral in Equation\,\eqref{eq:appendix_dispersion_first_line}, we convert to polar coordinates in $\boldsymbol{s}$ with new variables $(s, \theta_s)$, where $\theta_s$ is chosen without loss to be the angle between $\boldsymbol{s}$ and $\boldsymbol{k}$. This means that $\boldsymbol{\hat{s}}=\text{cos}(\theta_s)\boldsymbol{\hat{k}}+\text{sin}(\theta_s)\boldsymbol{\hat{k}_\perp}$, where $\boldsymbol{\hat{k}_\perp}$ is the unit vector perpendicular to $\boldsymbol{\hat{k}}$. The integral then becomes
\begin{equation}
\begin{split}
    &\int_{\mathbb{R}^2} \boldsymbol{\hat{s}}\,\Omega\left(\frac{s}{\xi}\right)\, e^{i\boldsymbol{k}\cdot\boldsymbol{s}} \diff^2 \boldsymbol{s} = \int_{0}^{\infty}\int_0^{2\pi}\left(\text{cos}(\theta_s)\boldsymbol{\hat{k}}+\text{sin}(\theta_s)\boldsymbol{\hat{k}_\perp}\right)\,\Omega\left(\frac{s}{\xi}\right)\,e^{iks\text{cos}(\theta_s)} s\diff s \diff \theta_s \\
    & = \int_{0}^{\infty}\Omega\left(\frac{s}{\xi}\right)\,s \left[\boldsymbol{\hat{k}}\int_0^{2\pi}\text{cos}(\theta_s)e^{iks\text{cos}(\theta_s)}\diff \theta_s + \boldsymbol{\hat{k}_\perp}\int_0^{2\pi}\text{sin}(\theta_s)e^{iks\text{cos}(\theta_s)}\diff \theta_s\right]\diff s \,.
\end{split}
\label{eq:appendix_s_hat_double_integral}
\end{equation}
Using the change of variables $\theta_s = \theta'-\pi$, we can see that the second angular integral evaluates to zero:
\begin{equation}
\begin{split}
    \boldsymbol{\hat{k}_\perp}\int_0^{2\pi}\text{sin}(\theta_s)e^{iks\text{cos}(\theta_s)}\diff \theta_s &  = \boldsymbol{\hat{k}_\perp}\int_{-\pi}^{\pi}-\text{sin}(\theta')e^{-iks\text{cos}(\theta')}\diff \theta' = 0 \,,
\end{split}
\end{equation}
where the second equality arises from the integrand being an anti-symmetric function integrated across symmetric limits. Equation \eqref{eq:appendix_s_hat_double_integral} thus simplifies to
\begin{equation}
    \begin{split}
    \int_{\mathbb{R}^2} \boldsymbol{\hat{s}}\,\Omega\left(\frac{s}{\xi}\right)\, e^{i\boldsymbol{k}\cdot\boldsymbol{s}} \diff^2 \boldsymbol{s} &=\int_{0}^{\infty}\Omega\left(\frac{s}{\xi}\right)\,s \left[\boldsymbol{\hat{k}}\int_0^{2\pi}\text{cos}(\theta_s)e^{iks\text{cos}(\theta_s)}\diff \theta_s \right]\diff s \\
    &=\boldsymbol{\hat{k}}\int_{0}^{\infty}\Omega\left(\frac{s}{\xi}\right)\,s \,\left[-i\frac{\partial}{\partial (ks)}\int_0^{2\pi}e^{iks\text{cos}(\theta_s)}\diff \theta_s\right]\diff s \\
    & = \boldsymbol{\hat{k}}\int_{0}^{\infty}\Omega\left(\frac{s}{\xi}\right)\,s \,\left[-i\frac{\partial}{\partial (ks)}2\pi J_0(ks)\right]\diff s \\
    & = \boldsymbol{\hat{k}}\int_{0}^{\infty}\Omega\left(\frac{s}{\xi}\right)\,s \,\left[i\, 2\pi J_1(ks)\right]\diff s,
    \end{split}
\end{equation}
where in the third line we used the standard integral form of the $0^\text{th}$ order Bessel function of the first kind, $J_0$, and in the fourth line we used the fact that $\frac{\diff}{\diff x}J_0(x)=-J_1(x)$. Substituting this result back into the dispersion relation yields
\begin{equation}
        \lambda(\boldsymbol{k}) = -k^2 + 2\pi\,\Runner \changes{\phi}(\Runner)\frac{\partial g}{\partial \runner}\Big\rvert_{\Runner}\,\frac{\mu}{\xi^2}\,k\,(\boldsymbol{\hat{k}}\cdot \boldsymbol{\underline{\underline{R}}}(\alpha) \boldsymbol{\hat{k}})\int_0^\infty \Omega\left(\frac{s}{\xi}\right)\, s\,J_1(ks) \diff s,
\end{equation}
which is the second line of Equation\,\eqref{eq:dispersion_single_species_direct}, as required. For further discussion of the role of Bessel functions and Hankel transforms in the linear stability analysis of these models, see Section 3 of \citet{jewell2023}.}

\section{\changes{Kernel assumptions}}
\label{sec:appendix_kernel_assumptions}

\changes{In order for $\Lambda_{uv}(k)$ to share the same sign as $\mu_{uv}$ for values of $k$ that are `relevant for pattern formation' -- typically those corresponding to the largest positive values of $\lambda(k)$, we require that: (a) $\changes{\phi}_u(\Chaser,\Runner)\frac{\partial g_{uv}}{\partial v}\Big\rvert_{V}>0$, which follows from our core model assumptions, and (b) the Hankel transforms of the interaction kernels are positive at these key values of $k$.}

\changes{Condition (b) automatically holds for any kernel whose Hankel transform is always positive. \citet{cho2024positivity_hankel} outline some \textit{sufficiency} conditions for this. For the direct sensing model, the primary condition is that $\sqrt{s}\Omega\left(\frac{s}{\xi}\right)$ is non-negative and decreasing. One way of satisfying this condition is by making a `splattergun' assumption: as we increase distance, $s$, then in 2D space, interactions are spread out over a circle of increasingly large circumference $2\pi s$, and so we might expect the interaction kernel to decay at least as fast as $\frac{1}{s}$, analogously to how light intensity and some physical forces decay as $\frac{1}{s^2}$ in 3D space. For the gradient sensing model, in order for the $0^\text{th}$ order Hankel transform to always be positive, there is the same condition that $\sqrt{s}\tilde{\Omega}\left(\frac{s}{\xi}\right)$ is non-negative and decreasing, as well as an additional requirement that $-\sqrt{s}\frac{\diff}{\diff s}\tilde{\Omega}\left(\frac{s}{\xi}\right)$ is non-negative and decreasing. Biologically speaking, all of the above conditions correspond to interactions that do not change direction with distance, and get sufficiently weaker as distance increases. \citet{cho2024positivity_hankel} also stipulate that certain integrals of the kernel remain finite and that the kernel vanishes at infinity, both of which are naturally satisfied in biological contexts.}

\changes{The above conditions are sufficient but not necessary, and many commonly studied kernels that do not satisfy them still have Hankel transforms that are always positive. In direct sensing models, for instance, both the exponential kernel
and the Gaussian-derivative kernel, $\Omega\left(\frac{s}{\xi}\right)\propto \frac{s}{\xi} e^{-0.5(\frac{s}{\xi})^2}$, used by \citet{Painter2015nonlocal, jewell2023, yurij_2025_nonlocal}, have positive $1^\text{st}$ order Hankel transforms. Similarly, for gradient sensing models, the exponential kernel and the Gaussian kernel, $\tilde{\Omega}\left(\frac{s}{\xi}\right)\propto e^{-0.5(\frac{s}{\xi})^2}$, used by \citet{potts2024_heterogeneous, fagan2017_exponential_gaussian_kernel, yurij_2023_review}, both have positive $0^\text{th}$ order Hankel transforms. Even in the case of the tophat kernel, where the Hankel transform can be negative for some values of $k$, the values of $k$ for which this occurs often do not correspond to those with the highest values of $\lambda(k)$, as discussed by \citet{jewell2023}. Essentially, the negative values of the transform have smaller magnitude than the positive values, and also occur at higher values of $k$, where the diffusion term, $-k^2$, is larger, further reducing $\lambda$.}

\changes{All of this is to say, that in a biological context, condition (b), and its corollary that $\text{sgn}(\Lambda_{uv}(k))=\text{sgn}(\mu_{uv})$ for relevant values of $k$, are both reasonable assumptions. These assumptions are useful for understanding the \textit{typical} effect of chase-and-run dynamics on pattern formation. Whilst relaxing these assumptions allows for a richer variety of phenomena, exploring them lies outside the scope of this work and is covered in \citet{jewell2023} for non-chiral interactions.}

\end{appendices}

\bibliography{sn-bibliography}

\end{document}